
\input harvmac

\def\tm{{\tilde m}}
\def\rone{{r_1}}
\def\rtwo{{r_2}}
\Title{PUPT--1276}
{{\vbox {\centerline{World Sheet and Space Time Physics}
\bigskip
\centerline{in Two Dimensional (Super) String Theory}
}}}

\centerline{P. Di Francesco}
\smallskip
\centerline{and}
\smallskip
\centerline{D. Kutasov}
\bigskip\centerline
{\it Joseph Henry Laboratories,}
\centerline{\it Princeton University,}
\centerline{\it Princeton, NJ 08544.}

\vskip .2in

\noindent
We show that tree level ``resonant'' $N$ tachyon scattering amplitudes,
which define a sensible ``bulk'' S -- matrix
in critical (super) string theory in any dimension,
have a simple structure
in two dimensional space time,
due to partial decoupling of a certain infinite
set of discrete states. We also argue that the general (non resonant)
amplitudes are determined by the resonant ones, and calculate them
explicitly, finding an interesting analytic structure. Finally,
we discuss the space time interpretation of our results.

\Date{9/91}
%

\newsec{Introduction.}

String theory
\ref\GSW{M. Green, J. Schwarz and E. Witten, Superstring theory
(Cambridge Univ. Press, Cambridge, 1987) and references therein.}\
is a prime candidate for a unified quantum
description of short distance physics, which naturally gives rise to
space-time gravity as well as gauge fields and matter.
However, our understanding of this theory is hindered by its
complexity, related to the enormous number of space-time degrees of
freedom (massive resonances), the proliferation of vacua, and lack of an
organizing (non-perturbative) dynamical principle.

In this situation, one is motivated to look for toy models which capture
some of the important properties of strings, while allowing for a more
complete understanding. In the last year important progress was made in
treating such toy models, corresponding to strings propagating
in a two dimensional (2D) space-time. The low space-time dimension drastically
reduces the number of degrees of freedom, eliminating most of the
massive oscillation modes of the string and leaving behind essentially
only the center of mass of the string (the `tachyon' field) as a physical
field theoretic degree of freedom.
Following the seminal work of
\ref\SURF{V. Kazakov, Phys. Lett. {\bf 150B} (1985) 282; F. David, Nucl.
Phys. {\bf B257} (1985) 45; J. Ambjorn, B. Durhuus and J. Frohlich,
Nucl. Phys. {\bf B257} (1985) 433; V. Kazakov, I. Kostov and A. Migdal,
Phys. Lett. {\bf 157B} (1985) 295.}
\ref\ORIG{D. Gross and A. Migdal, Phys. Rev. Lett. {\bf 64}
(1990) 127; M. Douglas and S. Shenker, Nucl. Phys. {\bf B335} (1990) 635;
E. Brezin and V. Kazakov, Phys. Lett. {\bf 236B} (1990) 144.}
\ref\CONE{
E. Brezin, V. Kazakov, and Al. Zamolodchikov, Nucl. Phys. {\bf B338}
(1990) 673; D. Gross and N. Miljkovi\'c, Phys. Lett {\bf 238B} (1990) 217;
P. Ginsparg and J. Zinn-Justin, Phys. Lett {\bf 240B} (1990) 333;
G. Parisi, Phys. Lett. {\bf 238B} (1990) 209.},
it was understood that the center of mass in these 2D string
theories is described by free fermion quantum mechanics
\ref\BDSS{T. Banks, M. Douglas, N. Seiberg and S. Shenker,
Phys. Lett. {\bf 238B} (1990) 279.}
\ref\SFT{S. Das and A. Jevicki, Mod. Phys. Lett. {\bf A5} (1990) 1639.}
\ref\SENG{A. Sengupta and S. Wadia,
Tata preprint TIFR/TH/90-13 (1990).}
\ref\GK{D. Gross and I. Klebanov, Nucl. Phys. {\bf B352} (1991) 671.}.
This remarkable phenomenon has led to rapid progress in the qualitative
and quantitative understanding of these theories
\ref\D{M. Douglas, Phys. Lett. {\bf 238B} (1990) 176.}
\ref\DIFK{P. Di Francesco and D. Kutasov, Nucl. Phys. {\bf B342} (1990) 589;
Princeton preprint PUPT-1206 (1990).}
\ref\MOORE{G. Moore,
Rutgers preprint RU-91-12 (1991).}
\ref\INTER{D. Gross and I. Klebanov, Princeton preprint PUPT-1241 (1991);
G. Mandal, A. Sengupta, and S. Wadia, IAS preprint, IASSNS-HEP/91/8 (1991);
K. Demeterfi, A. Jevicki, and J.P. Rodrigues, Brown preprints BROWN-HET-795,
803
(1991); J. Polchinski, Texas preprint UTTG-06-91 (1991).}.

This progress was phrased in the language of matrix models of random surfaces
\ref\MATMOD{
E. Brezin, C. Itzykson, G. Parisi and J.B. Zuber, Comm. Math. Phys.
{\bf 59} (1978) 35 ; D. Bessis, Comm. Math. Phys. {\bf 69} (1979) 147};
it is important to understand the results and in particular the free fermion
structure in the more familiar Polyakov path integral formulation of
2D gravity
\ref\POL{A. Polyakov, Phys. Lett. {\bf B103} (1981) 207, 211.}.
If we are to utilize the impressive results of 2D string theory
in more physically interesting situations, which are either hard to
describe by means of matrix models (e.g. fermionic string theories)
or can be described by matrix models which are hard to solve (e.g.
$D>2$ string theories), we must learn how to handle the continuum
(Liouville) theory
more efficiently. Despite important progress in this direction
\ref\CT{T. Curtright and C. Thorn, Phys. Rev. Lett. {\bf 48} (1982) 1309;
E. Braaten, T. Curtright and C. Thorn, Phys. Lett. {\bf 118B} (1982) 115;
Ann. Phys. {\bf 147} (1983) 365.}
\ref\GN{J.L Gervais and A. Neveu, Nucl. Phys. {\bf B209} (1982) 125;
{\bf B224} (1983) 329; {\bf B238} (1984) 125.}
\ref\NATI{N. Seiberg, Rutgers preprint RU-90-29 (1990).}
\ref\JP{J. Polchinski, Texas preprints UTTG-19-90; UTTG-39-90 (1990).}
some aspects of the matrix model results are still mysterious.

The purpose of this paper is to try and probe the continuum string theory
in various ways, with the hope of understanding the underlying free fermion
structure. We will not be able to get as far as that, but we will
see aspects of the simplicity emerging.
Most of our analysis will be done on the sphere; the matrix
model techniques are (so far) much more powerful
in obtaining higher genus results. As a compensation,
the spherical structure will be quite well understood;
in fact, many of the results described below were not
obtained from matrix models (so far).

What can we hope to learn from such an endeavour?
The free fermion structure of 2D string theory is highly unlikely to survive
in more physically interesting situations. However, there are some features
which are expected to survive: the $(2g)!$ growth of the perturbation
expansion
is expected
\ref\SHENKER{S. Shenker, Rutgers preprint RU-90-47 (1990).}\
to be a generic property of all
(super--) string theories; issues related to background independence of the
string field theory, the form of the (classical) non linear
equations of motion in string field theory,
and even the right variables in terms of which one should
formulate the theory may be studied in this simpler context.
The advantage of such a simple solvable framework is to provide a laboratory
to quantitatively check ideas in string field theory.
The fact that we do not quite understand the matrix model results from the
continuum is significant: it suggests that a new point of view on the
existing techniques or new techniques are needed for treating strings.
Finally, it was argued recently
\ref\WBH{E. Witten, IAS preprint IASSNS-91/12 (1991).}\
that one can study space-time singularities in string
theory using related two dimensional string models.
Issues related to gravitational back reaction can
be naturally described and studied in the continuum approach.

The paper is organized as follows.
In section 2, after an exposition of tachyon propagation in $D$ dimensional
string theory,
we discuss
in detail 2D bosonic strings, or more precisely $c \leq 1$
Conformal Field Theory (CFT)
coupled to gravity. In the conformal gauge we are led
to study (minimal or $c=1$) matter with action ${\cal S}_M (g)$
(on a Riemann surface with metric $g$), coupled to the Liouville mode.
The action is \POL:
\eqn\a{ {\cal S} ={\cal S}_L({\hat g}) + {\cal S}_M({\hat g}) }
where the dynamical metric is $g_{ab} = e^{\phi} {\hat g}_{ab} $ and:
\eqn\SL{ {\cal S}_L({\hat g})={1 \over {2 \pi}} \int \sqrt{\hat g}[
{\hat g}^{ab} \partial_a \phi \partial_b \phi -{Q \over 4}{\hat R} \phi
+2 \mu e^{\alpha_+ \phi}] }
with $Q$ and $\alpha_+$ finitely renormalized parameters
\ref\DDK{F. David, Mod. Phys. Lett. {\bf A3} (1988) 1651;
J. Distler and H. Kawai, Nucl. Phys. {\bf B321} (1989) 509.} (see below).
It is very useful to think about the Liouville mode as a target space
coordinate,
and of \a\ as a critical string system in a non-trivial
background
\ref\IND{S. Das, S. Naik and S. Wadia, Mod. Phys. Lett. {\bf A4} (1989)
1033; J. Polchinski, Nucl. Phys. {\bf B324} (1989) 123;
T. Banks and J. Lykken, Nucl. Phys. {\bf B331} (1990) 173.}.
This point of view proves helpful for the analysis of the
Liouville dynamics \NATI, which is given by a non-trivial interacting
CFT \SL. The exponential interaction in \SL\ keeps the Liouville field
away from the region where the string coupling $g_{st} = g_0 e^{-{Q \over 2}
\phi}$ blows up ($\phi \to -\infty$ in our conventions).
Due to the presence of this `Liouville wall', correlation functions in this
theory are non-trivial. To understand them, it is useful to break up the
problem into two parts.
It is clear that studying the scattering {\it in the bulk} of the
$\phi$ volume is a simpler task than that of considering the general
scattering processes.
Since such bulk amplitudes are insensitive to the precise form
of the wall (as we'll
explicitly see
later), they can be calculated
using free field techniques. This is the first step which is performed
in section 2.

The results for bulk amplitudes
are puzzling if one compares them with the well known structures
arising in critical string theory. There, bulk scattering is the only effect
present, and it is described by a highly non-trivial S -- matrix, incorporating
duality, an infinite number of massive
resonances etc. The main differences between this situation and ours are:

\noindent{}1) The critical string amplitudes are meromorphic in the external
momenta. When the integral representation diverges, one calculates the
amplitudes by analytic continuation. We will see that in 2D string
theory the situation is more involved (this is expected to be a
general property of all ($D\not=26$) non critical string theories).

\noindent{}2) The bulk scattering amplitudes in the 2D problem exhibit
miraculous symmetries (first noticed in
\ref\PO{A. Polyakov, Mod. Phys. Lett. {\bf A6} (1991) 635.}).
Most of the tachyon scattering amplitudes vanish. Those which do not, have
an extremely simple form which is strongly reminiscent of the corresponding
free fermion expressions \MOORE, \INTER. These phenomena
are far from completely
understood, and have to do on the one hand with the small
number of states and large symmetry in the theory,
and on the other with peculiarities
of (massless) 2D kinematics.
For all $D>2$, the form of the amplitudes is qualitatively similar to
that in the critical case $D=26$. Hence, an abrupt change in the
behavior of the theory occurs between $D=2$ and $D>2$.

At the second stage,
after treating tachyon scattering in the bulk, we proceed and consider
the generic scattering amplitudes which probe the structure of the Liouville
wall.
A direct approach seems unfeasible and we argue instead that one can deduce the
general structure of the interactions from their bulk part. The main idea
is that the Liouville interaction \SL\ represents (in target space language)
a tachyon condensate. If we understand the interaction of tachyons in the
bulk, it is reasonable to expect that we can understand the dynamical effect
of the wall. This procedure is nevertheless not guaranteed to work apriori,
but it does here (in $2D$),
and this allows us to obtain the full tree level tachyon
scattering matrix. The most remarkable feature of this
S -- matrix is that one can write down all $N$ point functions
{\it explicitly}.
Scattering amplitudes are naturally expressed in terms of target space
Feynman rules with an infinite number of calculable irreducible N particle
interactions, which can be thought of as arising from integrating
out the massive (discrete) modes. One of the main technical results of this
section is the evaluation of these irreducible vertices.
We also discuss the space-time picture which emerges from this treatment
of the tachyon, and apply the results
to calculations of correlation functions in minimal models
\ref\BPZ{A. Belavin, A. Polyakov and A. Zamolodchikov, Nucl. Phys.
{\bf B241}, 333 (1984).}
coupled to gravity, reproducing the results
of the KdV formalism
\D, \DIFK.

In section 3 we apply the techniques of section 2 to the problem of
calculating correlation functions in fermionic string theory
(again in 2D).
As expected from general arguments, there is little qualitative difference
between this case and the simpler bosonic one.
The only field theoretic degree of freedom in the Neveu-Schwarz (NS)
sector is again the
massless ``tachyon'' (the center of mass of the string); the Ramond (R)
sector contains an additional (massless)
bosonic space-time field.
We find that the massless
sector scattering picture in fermionic 2D string theory
is similar to the one obtained in the bosonic case.
The only difference is in the spectrum of discrete
states in the two models; the way it affects the scattering illuminates
the role of the latter.
We mention the possibility
\ref\KUS{D. Kutasov and N. Seiberg, Phys. Lett. {\bf 251B} (1990) 67.}
of obtaining stable (tachyon free)
superstring theories at $D\geq 2$ by a chiral GSO projection
of the fermionic string, and show that the 2D superstring is
topological.

Section 4 contains some comments on the physics of discrete
massive states in
2D string theory.
Those are important from several points of view. First, they represent
the only remnants of the infinite tower of
massive states -- the hallmark of
string theory -- and it would be interesting to study their dynamics.
Second, these discrete operators are instrumental to the question
of gravitational back reaction in two dimensional string theory
\WBH,
\ref\BER{M. Bershadsky and D. Kutasov, Princeton preprint PUPT-1261, (1991).},
and by understanding their dynamics we may study issues
related to gravitational singularities in string theory.
Finally they are closely related to the large symmetry of 2D
string theory.

Section 5 contains some summarizing remarks.
In appendices A,B we compare Liouville results with those of matrix models
(given by generalized KdV equations \D\ for minimal models) and describe some
features of the 1PI tachyon amplitudes.

\bigskip

\newsec{Tachyon Dynamics in Bosonic String Theory.}

\subsec{The general structure and strategy.}

We will concentrate throughout this paper on the situation in
string theory in two dimensional space time,
where many special features arise.
It will be very useful to have in mind the perspective of the higher
dimensional
situation for comparison. We will describe it in this subsection, in addition
to defining some concepts which will be useful
later, and describing the procedure
which we will use to calculate the S -- matrix.

Thus we start with the Polyakov string in flat $d$ dimensional (Euclidean)
space
\eqn\polst{{\cal S}_M(X,g)=
{1\over2\pi}\int\sqrt{g}g^{ab}\partial_a X^i\partial_b X^i}
$i=1...d$. The most convenient prescription \POL\ to quantize
this generally covariant two dimensional system is to fix a conformal gauge
$g_{ab}=e^\phi\hat{g}_{ab}$, in which the system is described by the
Liouville mode $\phi$ and space coordinates $X^i$,
living in the background metric $\hat g$
(the gauge fixing also introduces reparametrization ghosts
$b,c$ with spins $2, -1$ respectively).
The action for the system is \a\ where the Liouville mode is governed
by \SL\ and the matter fields $X^i$ by the free scalar action
\polst\ with $g\rightarrow\hat g$, the non dynamical background (``fiducial'')
metric
\foot{We don't want to leave the impression that the equivalence
of \polst\ and \a, \SL\ is well understood. There are subtleties
related to the measure of $\phi$ \POL, \DDK\ and the conformal
invariance of \SL. Our point of view is that \SL\
defines a CFT (in a specific regularization to be discussed below),
so that we are certainly studying a consistent background of
critical string theory. The world sheet physics obtained is also
reasonable, thus it is probably the right quantization of
$2d$ gravity. The relation of $\phi$ in \SL\ to the conformal
factor of $g_{ab}$ is at best a loose one.}.
The parameters in \SL\ are determined by requiring gauge invariance
(independence of the arbitrary choice of $\hat g$). This is equivalent \DDK\
to BRST invariance with
$Q_{BRST}=\oint cT$, ($T=T_L+T_M$ is the total stress tensor of the system),
which fixes
\eqn\QA{Q=\sqrt{25-d\over3};\;\alpha_+=-{Q\over2}+\sqrt{1-d\over12}}
{}From the critical string point of view, BRST invariance is the requirement
that the matter $+$ Liouville system be a consistent background of the
$D=d+1$ dimensional critical
bosonic string. Thus it is superficially very similar to ``compactified''
critical string theory, where one also replaces part of the matter system
by an arbitrary CFT with the same central charge (here the Liouville CFT).
The most important difference is that the density of states of the string
theory
is {\it not} reduced by compactification, while it is reduced by Liouville.
In other words, although the central charge of the Liouville theory
\eqn\CL{c_L=1+3Q^2}
is in general larger than one, the density of states
is that of a $c=1$ system (see
\NATI,
\ref\NADA{D. Kutasov and N. Seiberg, Nucl. Phys. {\bf B358} (1991) 600.}
for further discussion).

We will concentrate on the dynamics of the center of mass of the string, the
tachyon field. Of course, for generic $D$ there is no reason to focus
on the tachyon, both because it is merely the lowest lying state
of the infinite string spectrum, and because it is tachyonic, thus absent in
more physical theories. Our justification will come later, when we'll
consider the two dimensional situation, where the tachyon
is the only field theoretic degree of freedom, and is massless
(we will still call it ``tachyon'' then).
The on shell form of the tachyon vertex operator is
\eqn\tach{T_k=\exp(ik\cdot X+\beta(k)\phi)}
where $k, X$ are $d$ -- vectors, and BRST invariance implies
\eqn\mass{{1\over2}k^2-{1\over2}\beta(\beta+Q)=1}
As in critical string theory ($D=d+1=26$), this equation is simply
the tachyon mass shell condition; the vertex operator $T_k$
is related to the wave function $\Psi$ of the
corresponding state through $T_k=g_{st}\Psi$
so that the wave function has the form
(recall $g_{st}\propto \exp(-{Q\over2}\phi)$)
\eqn\ps{\Psi(X,\phi)=\exp\left(ik\cdot X+(\beta(k)+{Q\over2})\phi\right)}
We thus recognize the Liouville momentum (or energy, interpreting Liouville
as Euclidean time)
$E=\beta+{Q\over2}$, and
space momentum $p=k$. Eq. \mass\ can be rewritten as
\eqn\tachmass{E^2=p^2+m^2;\; m^2={2-D\over12}}
reproducing the well known value of the ground state
energy of $D$
dimensional strings.

{}From the world sheet point of view \NATI, \JP, the region $\phi\rightarrow
\infty$ corresponds to small geometries in the dynamical metric $g$
\polst. This is also the region where the string coupling constant
$g_{st}\rightarrow0$ and the Liouville interaction in \SL\ is
negligible. From eq. \mass\ we see that on shell states fall into
three classes \NATI, \JP:

\noindent{}1) $E=\beta+{Q\over2}>0$: the wave
function $\Psi$ \ps\ is infinitely
peaked at small geometries (in the dynamical metric $g$)
$\phi\rightarrow\infty$. Insertion of such operators
into a correlation function corresponds to local disturbances of the surface.

\noindent{}2) $E<0$: the wave function is infinitely peaked at
$\phi\rightarrow-\infty$. Such operators do not correspond
to local disturbances of the surface. In \NATI, \JP\ it was argued
that they do not exist.

\noindent{}3) $E$ imaginary: $\Psi(X,\phi)$ is in this case
($\delta$ function) normalizable. Such states create finite holes
in the surface and destroy it if added to the action. Thus they correspond
to world sheet instabilities. In space time, such operators correspond
to tachyonic string states (real Euclidean $D$ momentum).
It is well known that one can at best make sense of theories
with tachyons on the sphere; at higher genus, on shell tachyons
in the loops cause IR divergences. The existence of such states in a string
theory is in one to one correspondence with existence of a non trivial
number of states \NATI, \NADA. The cosmological operator in \SL\
corresponds to a macroscopic state for $d>1$ \QA.

The main object of interest to us here will be the tachyon S -- matrix, the set
of amplitudes\foot{We will be sloppy with integral signs. In $N$ point
functions $N-3$ of the vertices should be integrated over.}:
\eqn\tachcor{A(k_1,..k_N)=\langle T_{k_1}..T_{k_N}\rangle}
where the average is performed with the action \a. Translational
invariance in $X$ implies momentum conservation
\eqn\momen{\sum_{i=1}^Nk_i=0}
There is no momentum conservation in the $\phi$ direction due to the
interaction, therefore in general all amplitudes \tachcor\ satisfying
\momen\ are non vanishing. The Liouville path integral is complicated, but some
preliminary intuition can be gained by integrating out the zero mode of
$\phi$, $\phi_0$
\ref\GTW{ A. Gupta, S. Trivedi and M. Wise, Nucl. Phys. {\bf B340}
(1990) 475.}.
Splitting $\phi=\phi_0+\tilde\phi$, where
$\int\tilde\phi=0$ and integrating in \tachcor\ $\int_{-\infty}^\infty
d\phi_0$,
we find:
\eqn\newA{A(k_1,..,k_N)=\left({\mu\over\pi}\right)^s\Gamma(-s)\langle
T_{k_1}... T_{k_N}
\left[\int\exp\left(\alpha_+\phi\right)\right]^s\rangle_{\mu
=0}}
In \newA\ the average is understood to exclude $\phi_0$
(and we have absorbed a constant, $\alpha_+$
into the definition of the path integral); note also that
it is performed with the free action \a, \SL\ : $\mu=0$. $s$ is the KPZ
\ref\KPZ{
V. Knizhnik,
A. Polyakov and A. Zamolodchikov, Mod. Phys. Lett {\bf A3} (1988) 819.}
\DDK\
scaling exponent:
\eqn\Spar{\sum_{i=1}^N\beta(k_i)+\alpha_+s=-Q}
The original non linearity manifests itself in \newA\ through
the (in general non integer) power of the interaction.

We seem to have gained nothing since for generic momenta
$s$ is an arbitrary complex number, and \newA\ is only a formal expression.
However now the space time interpretation is slightly clearer.
Amplitudes with $s>0$ (assume $s$ real for simplicity) are
dominated by the region $\phi\rightarrow\infty$ in the zero mode integral
(the region far from the Liouville wall); those with $s<0$ receive their main
contribution from the vicinity of the wall.
As $s\rightarrow0$ we see an apparent
divergence in \newA\ (more generally this happens whenever $s\in Z_+$).
{}From the world sheet point of view this is a trivial effect;
the Laplace transformed amplitude is finite everywhere:
\eqn\Lap{\mu^s\Gamma(-s)=\int_0^\infty dAA^{-s-1}\exp(-\mu A)}
{}From \Lap\ we see that the $s\rightarrow0$ divergence at fixed $\mu$
is a small area divergence in the integral over areas $A$.
{}From this point of view the right way to interpret \newA\
for $s\in Z_+$ is to replace
$\mu^s\Gamma(-s)\rightarrow{(-\mu)^s\over s!}\log{1\over\mu}$.
This so called ``scaling violation'' is of course in perfect
agreement with KPZ scaling of the fixed area amplitudes. In space time
the picture is more interesting; at $s=0$ the amplitude balances itself
between being exponentially dominated by the boundaries of
$\phi$ space and receives contributions from
{\it the bulk} of the $\phi_0$ integral.
Thus such amplitudes represent scattering
processes that occur in the bulk of space time, and one would expect
them to be insensitive to the precise form of the wall, which from this
point of view is a boundary effect. That this is indeed the case is easily
seen in \newA. The coefficient of $\log\mu$ is given by
a free field amplitude -- the interaction disappears. Of course, it is
natural to interpret $\log{1\over\mu}$ as the volume of the Liouville
coordinate $\phi$ (remember that the wall
effectively enforces $\phi\geq\log\mu$, and one may
introduce a UV cutoff $\phi\leq\phi_{UV}$ \PO; the bulk amplitudes
per unit $\phi$ volume will be clearly independent
of $\phi_{UV}$, if the latter is large enough, as can be readily
verified by repeating the considerations leading to \newA).

Amplitudes with $s\in Z_+$
(more precisely the coefficients of
$\mu^s\log\mu$
or, equivalently, the fixed area
correlators at integer $s$) are also seen to simplify since they too
reduce to free field integrals \newA. The space time interpretation
is again clear -- these processes correspond to resonances of the scattering
particles with the wall -- the energy is precisely such that they can scatter
against $s$ zero momentum tachyons
(which are the building blocks of the ``wall'' \SL) in the bulk
of the $\phi$ volume. Of course, given all $s=0$ (bulk) amplitudes,
the general $s\in Z_+$ ones immediately follow by putting some
momenta to zero.

After understanding the nature of the difficulties
we're facing, we now turn to the strategy that we'll use to
obtain the amplitudes $A(k_1,..,k_N)$ \newA. We will proceed
in three stages:

\noindent{}$\underline{\rm Step\;\; 1:}$
Calculate \newA\ for $s\in Z_+$. For generic $D$ this step is
technically hopeless;
the analytic structure of the amplitudes is complicated
and it is not known how to perform the free field integrals
in \newA.
This is essentially due to the complicated
back reaction that occurs when tachyons propagate
in space time.
For $D=2$ two miracles occur:
first, the kinematics allows a finite region
in momentum space where the integral representation \newA\
converges, which is usually not the case for massless/massive
particles.
It is nice that such a region exists, since unlike critical
string theory, the amplitudes here
{\it can not} be continued analytically: they do not define
meromorphic functions of the momenta, because of
non conservation of Liouville momentum (energy), associated with the
existence of the exponential wall.
More importantly, we will be able
to actually calculate the integrals \newA\ in the above kinematic region,
and find simple results.
This will imply that the back reaction is much simpler (and milder)
in two dimensions than in general, and will allow us to recover the full
dynamical effect of the Liouville wall.

\noindent{}$\underline{\rm Step\;\; 2:}$
The result of the first step will be the function
$A(k_1, .., k_N)$ \tachcor\ for $s\in Z_+$ in the kinematic region
where the integral representation \newA\ converges.
The first remaining question
is how to calculate the general $N$ point functions \tachcor\
(with $s\not\in Z_+$)
in this kinematic region.
It is not known how to make sense of \newA\ in this general case.
One expects the qualitative behavior to be different in two
dimensions and in $D>2$. In the two dimensional case we will argue
that one can obtain the result by a physical argument.
We will see that the integer $s$ tachyon amplitudes are polynomials
in momenta (in an appropriate normalization). This
will be interpreted as the result of the fact that tachyon
dynamics can be described by a local two dimensional field theory
(obtained by integrating out the massive discrete string modes),
which for large momentum gives algebraic growth
of the amplitudes
(associated presumably
with a UV fixed point). The requirement that {\it all} amplitudes
must be polynomial in this normalization will fix them uniquely.
We would like to stress that the above argument is a phenomenological
observation which gives the right result; we do not know why the local
tachyon field theory appears.

\noindent{}$\underline{\rm Step\;\; 3:}$
After obtaining the amplitudes \tachcor\ for generic $s$ in
the region where
the integral representation converges, we will be faced with the last
problem: extending the results to all momenta $\{k_i\}$.
Recall that due to the non trivial background we can not
analytically continue. We will
see that from general Liouville considerations we expect
cuts in amplitudes and will suggest a physical picture
based on the above space time field theory, which
allows one to calculate all amplitudes. The integrals
over moduli space will be split to
contributions of intermediate tachyons (coming from regions
of degeneration), and an infinite sum over the discrete massive states,
which will give irreducible $N$ point vertices. The tachyon propagator
will be seen to be non analytic (containing cuts
at zero intermediate momenta), while the vertices
will be found to be analytic (in $
\{k_i\}$). We will give a general procedure for calculating these irreducible
1PI vertices.

The program described above can not be carried out
for $D>2$. We can understand the nature of the difficulties
and gain additional intuition by studying the $s=0$
four point function of tachyons, which
can be calculated for all $D$, as in the critical string case \GSW.
Thus we consider $A_{s=0}(k_1,..,k_4)$, which
is given using \newA, \Spar\ by:
\eqn\szero{A_{s=0}(k_1,..,k_4)=\pi\prod_{i=2}^4{\Gamma(k_1\cdot k_i-\beta_1
\beta_i+1)\over\Gamma(\beta_1\beta_i-k_1\cdot k_i)}}
The amplitude \szero\ exhibits an infinite set of poles
at
\eqn\poles{k_1\cdot k_i-\beta_1\beta_i+1=-n;\;n=0,1,2,...}
The meaning of these poles is clear; the $s=0$
amplitudes have the important property that
they conserve Liouville momentum, $\exp(\beta_1\phi)\exp(\beta_2\phi)
=\exp(\beta_1+\beta_2)\phi$, as opposed to the general
Liouville amplitudes that don't (due to the existence
of the Liouville wall) as explained above\foot{Note that Liouville
theory seems to exhibit the peculiar property that the OPE depends
on the particular correlation function considered (through $s$).};
this is of course
the reason why they are calculable. Thus the intermediate
momentum and energy in the $(1,2)$ channel, say, are
$k_{\rm int}=k_1+k_2$, $E_{\rm int}=\beta_1+\beta_2+Q/2$ (the shift
by $Q/2$ is as in \ps). The poles \poles\ occur when
\eqn\inter{E_{\rm int}^2-k_{\rm int}^2={2-D\over12}+2l}
Thus the poles in \szero\ correspond to on shell intermediate
tachyons ($l=0$), gravitons ($l=1$), etc
\foot{The graviton is only massless in $D=26$ \inter.}.
They carry the information about the non trivial back reaction
of the string to propagation of tachyons in space time. In world sheet
terms we learn that
trying to turn on a tachyon condensate in the action
spoils conformal invariance -- switches on a non zero $\beta$ function
(infinite correlation functions \szero\ signal logarithmic
divergences on the world sheet, as in dimensional regularization).
To restore conformal invariance we must correct the tachyon background
and turn on the other massless and massive string modes as well.

In space time terms, we conclude that the tachyon background
\tach\ while being a solution to the linearized equations
of motion of the string is not a solution to the
full non linear (classical) equations of motion and must be corrected,
both by correcting $T(X, \phi)$ and turning on the other modes
\ref\CFMP{C. Callan, D. Friedan, E. Martinec and M. Perry,
Nucl. Phys. {\bf B262} (1985) 593.}.
This is standard in string theory; we'll see later that while
the form \szero\ is still correct for $D=2$, the physical picture
is quite different.

For more than four particles, the $s=0$ amplitude \newA\ is given by
the usual Shapiro -- Virasoro integral representation \GSW:
\eqn\Npoint{A_{s=0}(k_1,..,k_N)=\prod_{i=4}^N\vert z_i\vert^{2(k_1\cdot
k_i-\beta_1\beta_i)}\vert 1-z_i\vert^{2(k_3\cdot k_i-\beta_3\beta_i)}
\prod_{4=i<j}^N\vert z_i-z_j\vert^{2(k_i\cdot k_j-\beta_i\beta_j)}}
No closed expression for \Npoint\ is known in general. The basic
problem in evaluating it is the complicated pole structure of
$A(k_1,..,k_N)$. There are many channels in which poles appear; to
analyze them quantitatively one has to consider the region of
the moduli integrals in \Npoint\ where some number of $z_i$ approach
each other. For example, to analyze the limit $z_4, z_5,.. ,z_{n+2}
\rightarrow0$, it is convenient to redefine
\eqn\zi{z_4=\epsilon,\;z_5=\epsilon y_5,\;..., z_{n+2}=\epsilon y_{n+2}}
and consider the contribution of the region $\vert\epsilon\vert<<1$
to \Npoint. By simple algebra we find an infinite number of poles at
$E={Q\over2}+\sum_i\beta_i,\;p=\sum_ik_i$ (sums over $i$
run over $i=1,4,5,6,..,n+2$) satisfying
$E^2-p^2={2-D\over12}+2l$ as in \inter. The residues of the poles
are related to correlation functions of on shell intermediate
string states. Indeed, by plugging \zi\ in \Npoint\ it is easy to
find the residues explicitly; for the first pole, e.g., we find
\eqn\resN{A_{s=0}(k_1,..,k_N)\simeq
{\langle T_{k_1}T_{k_4}...T_{k_{n+2}}T_{\tilde k}\rangle
\langle T_{\Sigma_i k_i}T_{k_2}T_{k_3}T_{k_{n+3}}.. T_{k_N}\rangle
\over ({Q\over2}+\sum_i\beta_i)^2
-(\sum_ik_i)^2-{2-D\over12}}}
where $\tilde k=-\sum_ik_i$.
The generalization of \resN\ for the higher poles is straightforward.
It is interesting that the amplitudes \Npoint\ have the standard space time
interpretation for all $D$.
Poles correspond to on shell intermediate states.
One can show decoupling of null states. The only special feature of $D=26$
is that in that dimension the vacuum that we are
considering is Lorentz invariant. We will use \Npoint\ to study the
dynamics of the theory.

Since the residues of the poles in \resN\ are in general non zero, we see that
$A(k_1,..,k_N)$ has many poles in all possible channels (corresponding
to different ways to cut the space time diagrams). This phenomenon
is a reflection of the complicated back reaction in string theory;
both the form of the space time equations of motion and their solutions
are untractable. Thus in the next subsections we'll turn to the situation
in $D=2$ where things are much simpler (but still very interesting).

\subsec{Two Dimensional String Theory and Minimal Models.}

\noindent{ \it 2.2.1. d $\leq$ 1 matter theories. }

In the rest of the section we'll be mainly interested in the theory
\polst\ with $D=d+1=2$, which consists of two scalar fields
$\phi, X^1=X $. It will be convenient to generalize slightly
by introducing a background charge for $X$:
\eqn\ma{
{\cal S}_M={ 1 \over {2 \pi}} \int \sqrt{ \hat g} \left[
{\hat g}^{ab} \partial_a X \partial_b X + {{i \alpha_0}
\over 2} {\hat R} X \right]}
Introducing $\alpha_0$ shifts
the central charge of the matter sector to:
\eqn\cc{ c=1-12 \alpha_0^2 \ \ \ ; \ \ \ \alpha_0 \in {\bf R} }
and furthermore has the effect of changing the momentum
conservation condition in \newA\ to
$\sum_{i=1}^N k_i=2\alpha_0(1-h)$ ($h$ -- genus). It is known
that in such cases we must insert certain screening charges
to make sense of the theory.

There are two main reasons to consider \ma. First,
this allows one to avoid considering zero momentum tachyons in the
action: from \tachmass\ we see that the cosmological term in
\SL\ has $E=0$ at $D=2$. We will encounter later subtleties
at $E=0$, thus it is convenient to shift $c$ as in \cc, in which case
we have in \tach\ $E=\beta+Q/2$, $p=k-\alpha_0$ and the on shell condition
\tachmass\ with $m^2=0$.
Following \NATI\ we choose the solution with positive $E$ (see
discussion in section 2.1):
\eqn\bet{\beta+{Q\over2}=\vert k-\alpha_0\vert}
We see that the tachyon is massless for all
$\alpha_0$, but $k=0$ does not
correspond to zero momentum ($p=0$) in general.
Thus $\alpha_0$ is a kind of IR regulator. The second
reason to study \ma\ is that for rational $\alpha_0^2$ one can restrict
the spectrum of $k$'s to a finite set of degenerate Virasoro representations;
this is the Feigin Fuchs construction
\ref\FF{B. Feigin and D. Fuchs, Funct. Anal. and Appl. {\bf 16}
(1982) 114.},
\ref\DF{V. Dotsenko and V. Fateev, Nucl. Phys. {\bf B240[FS12]} (1984) 312;
{\bf B251[FS13]} (1985) 691.}
of the BPZ minimal models \BPZ.

The conformal primaries are represented by vertex operators
$V_k = e^{ikX}$, with dimensions $\Delta_k = {1 \over 2}k(k-2 \alpha_0)$.
For the minimal models,
to evaluate the flat space CFT
correlation functions $\langle V_{k_1}...V_{k_N} \rangle$
one has to insert a number of screening operators of dimension $1$,
$V_{d_-}$, $V_{d_+}$,
integrated over the world sheet;
$d_{\pm}$ are the solutions of:
$${ {1 \over 2} d_\pm(d_\pm- 2 \alpha_0)=1, }$$
Momentum conservation implies
\eqn\neutcond{ \sum_{i=1}^N k_i + m d_- +n d_+ = 2 \alpha_0}
Although the structure for rational $\alpha_0^2$ is much richer than the
generic one,
it is easier to calculate correlators including
screening at irrational $\alpha_0^2$ and to analytically continue them
to rational $\alpha_0^2$ (see \DF\ for details).
Furthermore, we will find it convenient to consider generic $k$'s
(not only those corresponding to degenerate representations).
In the application to $c=1$ we are interested in correlators
with $n,m=0$ (and generic $k$). At the end of the calculation
we should take $\alpha_0\rightarrow0$.

What is the space time picture corresponding to string theory with matter
given by \ma? The action \a\  takes in this case the form:
\eqn\Srotated{\eqalign{
{\cal S}={ 1 \over {2 \pi}} \int [
&\partial X \bar\partial X + {{i \alpha_0}
\over 2} {\hat R} X+\lambda_+\exp(id_+X)+\lambda_-\exp(id_-X)\cr
+&\partial\phi\bar\partial\phi-{Q\over4}{\hat R}\phi+\mu\exp(\alpha_+\phi)
]\cr}}
Note the screening charges in the action. The $X$
zero mode integration enforces \neutcond. Naively \Srotated\
is related in a simple way to the $d=1$ system: by
redefining $\tilde\phi={Q\over2\sqrt2}\phi-{i\alpha_0\over\sqrt2}X$;
$\tilde X={Q\over2\sqrt2}X+{i\alpha_0\over\sqrt2}\phi$ we seem to find
in terms of $\tilde\phi,\tilde X$ a $d=1$ string in a background
given by \Srotated\ (expressed in terms of
$\tilde\phi,\tilde X$). We will see later that this is not
quite true, but qualitatively \Srotated\ still describes
(before restricting to the minimal models) a solution to $2D$
critical string theory, and its physics is very similar
to that of the $\alpha_0=0$ theory
(see also
\ref\MPY{D. Minic, J. Polchinski and Z. Yang, Texas
preprint UTTG-16-91 (1991).}).

\noindent{\it 2.2.2. Three point correlators without screening.}

We start by considering
the simplest case of bulk correlators of
three tachyons
without screening ($n=m=0$ in \neutcond), with $s$ zero momentum
tachyons (punctures).
Here we follow closely
\ref\DFK{P. Di Francesco and D. Kutasov, Phys. Lett.
{\bf261B} (1991) 385.}; this case will allow us to discuss some important
general features of the theory in a relatively simple context,
where the results of all the necessary intergals are known.
One has to evaluate \newA:
\eqn\ampiii{ A(k_1,k_2,k_3)=(- \pi)^3
\left({\mu\over\pi}\right)^s\Gamma(-s)\langle
T_{k_1}(0)T_{k_2}(\infty) T_{k_3}(1)
\left[\int\exp\left(\alpha_+\phi\right)\right]^s\rangle}
where we have used $SL(2,{\bf C})$ invariance to fix the positions of
the three tachyons
and redefined the path integral
by a factor of $(-\pi)^3$
for later convenience. The momenta are subject to the conservation laws:
$$ k_1+k_2+k_3 = 2 \alpha_0$$
\eqn\consl{ s \alpha_+ + \vert k_1-\alpha_0 \vert+\vert k_2 -\alpha_0 \vert
+\vert k_3- \alpha_0 \vert = {Q \over 2} }
With no loss of generality, we can take $k_1 \geq \alpha_0$,
$k_2 \geq \alpha_0$
and $k_3 \leq \alpha_0$.
The (bulk) amplitude \ampiii\ for integer $s$ can be expressed in terms of
known
integrals \DF\ (we introduce the notation $\Delta(x)\equiv\Gamma(x) /
\Gamma(1-x)$):
\eqn\threept{
\eqalign{ \langle T_{k_1} T_{k_2} T_{k_3} \left(\int e^{\alpha_+ \phi}
\right)^s \rangle =
\prod_{j=1}^s \int d^2w_j \vert w_j \vert^{2 \alpha}
\vert 1 - w_j \vert^{2 \beta} \prod_{1 \leq i<j \leq s}\vert w_i -w_j
\vert^{4 \rho} \ \ \ \ \ \ \ \ \cr
=(s!)(\pi \Delta(-\rho))^s \prod_{i=0}^{s-1} \Delta((i+1)\rho)
\Delta(1+\alpha+i\rho)\Delta(1+\beta+i\rho) \Delta(-1-\alpha-\beta-
(s+i-1)\rho) \cr}}
where we have performed the Wick contractions for the free fields
$X$ and $\phi$ using the propagators $\langle X(z) X(0) \rangle =
\langle \phi(z) \phi(0) \rangle = - \log \vert z\vert^2$, and:
\eqn\kinem{{ \alpha=-\alpha_+ \beta(k_1) \ \ \ ;\ \ \ \beta=-\alpha_+
\beta(k_3)
\ \ \ ;\ \ \ \rho=-{\alpha_+^2 \over 2} }}
The on shell kinematics \consl\ implies:
\eqn\p{\beta=\cases{\rho(1-s)&$\alpha_0>0$\cr
                    -1-\rho s&$\alpha_0<0$,\cr}}
Plugging \p\ in \threept, \ampiii\ we get (for $s\geq1$):
\eqn\resiii{\eqalign{\alpha_0>0:&\;\;A(k_1,k_2,k_3)=0\cr
\alpha_0<0:&\;\;A(k_1,k_2,k_3)=-\pi\Delta(-s)\left[\mu\Delta(-\rho)\right]^s
\prod_{i=1}^2(-\pi)\Delta(m_i)\cr}}
where
\eqn\mi{m_i={1\over2}\beta_i^2-{1\over2}k_i^2}
As discussed above, the apparent infinity due to $\Gamma(-s)$
is irrelevant at fixed area, and yields a logarithmic
correction at fixed $\mu$. In fact for $\alpha_0<0$, \p\ implies
that $m_3=-s$ so that we can rewrite $A$ \resiii\ as
\eqn\resthree{ A(k_1,k_2,k_3)= (\mu \Delta(-\rho))^s \prod_{i=1}^3 (- \pi
\Delta(m_i)) }
There are two puzzling features in \resiii:

\noindent{}1) We seem to find different results for the two
signs of $\alpha_0$; but from
\ma, \cc\ it is clear that physics must be independent of this sign.

\noindent{}2) The $\alpha_0\rightarrow0$ ($c\rightarrow1$) limit
is singular since $\Delta(-\rho)\rightarrow0$.

The resolution of these puzzles is quite instructive. We will see later
that \resthree\ is the general tachyon three point function. Then it is
clear that if nothing special happens, for integer $s=n$ we should
have $A(k_1,k_2,k_3)=\mu^n F(k_1,k_2,k_3)$ with some finite $F$. But this
is equivalent to a vanishing fixed area amplitude (see \Lap). In order to have
a non zero fixed area amplitude at $s\in Z_+$, $F_{s\rightarrow n}
(k_1,k_2,k_3)$
{\it must diverge}. The only difference between positive and negative
$\alpha_0$ is that for $\alpha_0>0$ all factors in \resthree\ are
finite (for generic $k$'s) while for $\alpha_0<0$,
$\Delta(m_3)=\Delta(-s)$ supplies the necessary divergence.
Eq. \resiii\ is an example of a general phenomenon: we will see later
that all bulk amplitudes vanish except those for which
all $k_i-\alpha_0$ {\it except one} have the same sign. Since we chose
$k_1,k_2>\alpha_0$, $k_3<\alpha_0$ and the $s$ punctures
in \threept\ correspond to $k-\alpha_0>0$ when $\alpha_0<0$
(and vice versa), \resiii\ is the natural result.
We see that the apparent discrepancy between positive and negative $\alpha_0$
in \resiii\ is due to the fact that we impose a ``resonance''
condition which is discontinuous.
Of course, although both signs of $\alpha_0$ are `right', it is more
useful to consider $\alpha_0<0$. This is what we'll do below.

The foregoing discussion seems to be at odds with our previous comments.
We have argued that if the integral representation \newA\
diverges, we can not continue analytically because of the expected appearance
of cuts in amplitudes. But of course for $\alpha_0>0$ the integral
representation is always divergent;
the simplest way to see that is to note that the integrand
in \threept\ is positive definite
while the integral \resiii\ is 0. Shouldn't we then discard the results in
this case?

A useful analogy is critical string tree level scattering. In that
case there is no range of momenta where the integral representation
converges for massless/massive particles, since the integral
representations for the different channels ($s,t,u$ for $N=4$) converge
in different, non overlapping kinematic regions, while the string
world sheet integral includes all channels. In that case one
splits the world sheet integral into several parts, calculates them
at different momenta and analytically continues, using the space time
picture as a guide to compute the divergent world sheet integral.
Here we do the same. That's why one can trust the divergent integral
\threept\ for $\alpha_0>0$. Liouville momentum is conserved
for bulk amplitudes; the key assumption is that there is a consistent
space time interpretation.

So far we have considered the first puzzle mentioned above. What happens
as we take $c\rightarrow1$? From \resthree\ we learn that the
operator
$\exp(\alpha_+\phi)=\exp(-{Q\over2}\phi)=\exp(-\sqrt{2}\phi)$
decouples in this case.
Notice that its wave function $\Psi$ \ps\ is constant, thus not peaked
at $\phi\rightarrow\infty$, and does not correspond to a local operator.
Its decoupling is consistent with \NATI, \JP.
However, there is another BRST invariant operator $\phi\exp(-{Q\over2}\phi)$
which is a candidate to play the role of the cosmological term (it is
known to be interesting
\ref\P{J. Polchinski, Nucl. Phys. {\bf B346}
(1990) 253.}).
Naively there seems to be a host of difficulties with this operator:
it is not
clear how to do the $\phi$ zero mode integral in \newA;
and also how to generalize the scaling arguments of \DDK\ to obtain
KPZ scaling, and the minisuperspace analysis of
\ref\MSS{G. Moore, N. Seiberg and M. Staudacher, Rutgers
preprint RU-91-11 (1991).}\ to get the Wheeler
de Witt equation (both KPZ scaling and the WdW equation are known to be valid
at $c=1$ from matrix models \CONE, \MOORE, \MSS). All these problems
are bypassed by turning on a small $\alpha_0$, and considering
the cosmological operator $V_c={1\over\Delta(-\rho)}\exp(\alpha_+\phi)$.
This operator has finite correlators as $c\rightarrow1$. It is indeed
equivalent to the previous one since for small $\epsilon=\alpha_++\sqrt2$,
$V_c\simeq{1\over\epsilon}\exp(-\sqrt{2}\phi)+\phi\exp(-\sqrt{2}\phi)$.
The leading divergent term vanishes inside correlation functions, as
remarked above. However,
in terms of $V_c$ all the above properties are manifest for all $c$;
the singularity at $c=1$ has been absorbed into an infinite
coupling constant renormalization (of $\mu$).

Our final result for the three point functions is \resthree. Remember
that it was obtained only for $s\in Z_+$ for the coefficient
of $\mu^s\log\mu$ and is equivalent to \resiii. This completes
step 1 in the program of section 2.1.

The amplitudes contain a product of ``wave function renormalization"
factors $-\pi\Delta(m_i)$ and it seems natural to define
`renormalized' operators
\eqn\redef{\tilde T_k={T_k\over(-\pi)\Delta({1\over2}\beta^2-{1\over2}k^2)}}
whose correlators
are much simpler. Applying \redef\ in
\resthree\ and defining $\mu$ as the coefficient
of $\tilde T_{k=0}$ (which also automatically implements the coupling
constant renormalization discussed above, since $\tilde T_{k=0}=V_c$)
we find:
\eqn\finalthree{\langle\tilde T_{k_1}\tilde T_{k_2}\tilde T_{k_3}
\rangle=\mu^s}
The second step now is to extend \finalthree\ to non integer $s$.
To do that we must use space time intuition. The main point is that
we find here and will see again for higher point functions
that correlation functions of $\tilde T$ \redef\ are polynomial in
momenta\foot{$\mu$
is irrelevant: it can be either put to 1 by a shift in $\phi$
or absorbed into the definition \redef\ of $\tilde T$ and the path
integral.},
for the cases where we can calculate them (in the bulk).
We would like to argue that this fact is an indication that tachyon
dynamics can be described by an effective local two dimensional field theory
obtained by integrating out the massive modes.
This is not usually the case in string theory; beyond low energy
approximations the light string states can not be described
(even classically)
by a local action. Tachyon amplitudes (e.g.) contain
poles corresponding to all the massive modes of the string \poles.
If we integrate out the latter we find a highly non local action.
In two dimensions the situation is better. The tachyon is the only
field theoretic degree of freedom. It interacts with an infinite
set of massive quantum mechanical degrees of freedom, which exist
only at particular (discrete) momenta. This interaction
is summarized by the normalization factors $\Delta(m_i)$ in
\resthree; space time gravity (and in general inclusion of the discrete
states) seems to have the mild effect
of renormalizing the tachyon field. The renormalized tachyon
$\tilde T$ is described by a 2D field theory. The fact
that its bulk three point function \finalthree\ is one, and more generally
that the bulk correlation functions obtained below are polynomial in momenta
is compatible with this suggestion. Thus we are led to postulate
that {\it all} correlators of $\tilde T$ must be polynomial in
external momenta\foot{In principle there could be tachyon poles in
amplitudes, but these would have shown up at integer $s$.
We will see later that they turn into cuts,
due to non conservation of energy.}. This will allow
us to fix them uniquely. E.g. for the three point function we conclude
that \finalthree\ is the general result for all $s$ (since the only
polynomial $P(k_i)$ which is 1 whenever $s\in Z_+$ is $P(k_i)=1$).

To recapitulate, two dimensional string theory has the striking property
that it is described by {\it two} consistent S -- matrices.
The one familiar from critical string theory is that for $T_k$ \tach, \tachcor.
It has poles corresponding to all on shell string states and
is crucial for the issue of the role of space time gravity in the theory;
we will return to it in section 4. However, in two dimensions
the role of space time gravity is mild; the renormalized field $\tilde T$
is described by a second S -- matrix, which follows from a two
dimensional field theory action. In fact, the action giving the set of
$\tilde T$ amplitudes is known from the matrix model approach
\SFT, \SENG, \GK.
In the rest of this section we will
describe in detail this S -- matrix. It is important to emphasize
that despite the simple relation \redef\ the two S -- matrices describe
genuinely different physics. For example, the $\tilde T$ S -- matrix
does not have bulk scattering
unlike that of $T$.
In fact, there is nothing special about amplitudes with $s=0$ at all
in this picture.
More importantly, gravitational physics is
absent in $\tilde T$. The new feature in two dimensional string theory
is that we seem to be able to turn off space time gravity!

The third step in the general program of section 2.1, involving the
extension of \finalthree\ to regions in $k_i$ where
the integral representation diverges is trivial here -- we do not expect
anything non trivial, since the full effect of Liouville momentum
non conservation is not felt in three point functions. It is
nevertheless instructive to examine the region of convergence
of \threept\ to make contact with the discussion of section 2.1.
{}From integrability as $w_i\rightarrow0,1$
in \kinem\ we find $\alpha+1>0, \beta+1>0$.
In terms of the Liouville momenta (and introducing
$\alpha_-=2/\alpha_+$, such that $-Q=\alpha_++\alpha_-$ and
for $\alpha_0<0$, $2\alpha_0=\alpha_-
-\alpha_+$ ) this implies a restriction on the energies:
\eqn\betak{\beta(k_i)>{\alpha_-\over2}=-{Q\over2}-{\alpha_+\over2}}
Since $\alpha_+$ is negative \QA\ we learn that the integral representation
for the correlators \threept\ only converges for states with
$E>{\vert\alpha_+\vert\over2}$ in agreement with the physical picture
presented in section 2.1
and with the discussion of
\NATI, \JP. Any state with $E>0$ can be treated by continuing
its correlation functions from $c\rightarrow-\infty$ ($\alpha_+\rightarrow0$).
The necessity to analytically continue in $c$ (or $\alpha_+$)
follows independently
in our approach from the precise
convergence conditions of \threept, which are nicely expressed in terms
of $m_i$ \mi: by \consl\ $m_1, m_2$ satisfy
$m_1+m_2=1+\rho s$, and the convergence conditions are $m_1, m_2>0$.
This is only consistent if $-\rho<{1\over s}$. Convergence of all $s$
amplitudes can only be achieved if $\rho\rightarrow0$ ($c\rightarrow-\infty$).
On the other hand for $E<0$ \threept\ is always divergent and one needs
additional space time physical input to understand this case.
This divergence is presumably related to the fact that for
$E<0$ the corresponding perturbation of the surface is not small.

The three
point functions for $c=1$ ($D=2$) string theory are thus given in complete
generality by \resthree, \finalthree\ (a $\delta(\sum k_i-2
\alpha_0)$ is understood
throughout). We will next consider the three point functions in minimal
models, for which we will have to introduce the screening charges $n,m$
in \neutcond.

\noindent{\it 2.2.3. Three point functions with screening
(minimal models).}

For minimal models, whose free field description was developed in \FF, \DF,
we should include arbitrary numbers of screening charges
$V_{d_\pm}$ in the `matter' amplitudes. As before, we will choose
$k_1, k_2>\alpha_0, k_3<\alpha_0$, which is necessary here to ensure that
two of the vertices are in one half of the Kac table, while the third
is in the other half \DF, and choose $\alpha_0<0$ for the same reasons
as before.
In this case we have $d_+=-\alpha_+, d_-=\alpha_-$. We are interested
in the result for $k$'s describing certain degenerate representations
and for rational $\alpha_+^2$, for which $s$ \Spar\ is in general
non integer. As before we will first tune $k_i, \alpha_+$ such that
$s$ is integer, and calculate
\eqn\threescr{
\eqalign{
A_{m,n}(k_1,k_2,k_3)=(-\pi)^3({\mu \over \pi})^s \Gamma(-s)
{\lambda_-^m \over m!}
\prod_{i=1}^m \int d^2t_i {\lambda_+^n \over n!} \prod_{j=1}^n
\int d^2 \tau_j \prod_{a=1}^s
\int d^2 w_a  \cr
\langle T_{k_1}(0) T_{k_2}(\infty) T_{k_3}(1)
\prod_{i=1}^m T_{\alpha_-}(t_i) \prod_{j=1}^n T_{-\alpha_+}(\tau_j)
\prod_{a=1}^s T_{0}( w_a) \rangle  \cr }}
The $t_i, \tau_i$ integrals over the locations of the screenings give
the matter correlation function; note the factors of $\lambda_+^n\over n!$,
$\lambda_-^m\over m!$ coming from expanding the action \Srotated. The
$w_a$ integrals come from Liouville.
The various 2D multiple integrals involved here have been computed \DF.
Due to the conservation laws:
\eqn\conscr{
\eqalign{ k_1 + k_2 + k_3 + m \alpha_- - n \alpha_+ &= 2 \alpha_0 \cr
\vert k_1 - \alpha_0 \vert + \vert k_2 - \alpha_0 \vert + \vert k_3
- \alpha_0 \vert + s \alpha_+ &= {Q \over 2} \cr } }
one obtains (after some algebra):
\eqn\rescr{ A_{m,n}(k_1,k_2,k_3)= (\mu \Delta(-\rho))^s (- \pi \Delta(-
\rho_+))^n (-\pi \Delta(-\rho_-))^m \prod_{i=1}^3 ( -\pi \Delta(m_i)) }
where $\rho_{\pm}= {\alpha_{\pm}^2 \over 2}=-m(\mp \alpha_{\pm})$.
The result \rescr\ is very similar to the case $n=m=0$ \resthree.
By adjusting the coeffients $\lambda_\pm$ of the screening charges
in \Srotated\ to $\lambda_\pm^{-1}=(-\pi)\Delta(-\rho_\pm)$
we can bring \rescr\ to the form \resthree. This choice of $\lambda_\pm$
is necessary already in the flat space CFT \DF. It is also very natural from
the point of view of \redef: in terms of $\tilde T$ the amplutdes $A_{m,n}$
are given again by \finalthree.
The continuation to non integer $s$ proceeds now in the same way as
for the case without screening, with the same conclusions.

Although \finalthree\ is our final result for the minimal model correlation
functions
(rational
$\alpha_0^2$), there is a slight subtlety in its interpretation. In that case
$k_i={1\over2}(1-r_i)\alpha_--{1\over2}(1-s_i)\alpha_+, i=1,2$,
$k_3={1\over2}(1+r_3)\alpha_--{1\over2}(1+s_3)\alpha_+$,
$\alpha_+^2={2p\over  p^\prime},p<p^\prime$, $r_ip^\prime>s_ip$
and $1\leq r_i\leq p-1,1\leq s_i\leq p^\prime-1$. Naively, the only fusion
rule for the minimal model three point functions \finalthree\
is \conscr, which is equivalent to
\eqn\fusion{r_1+r_2\geq r_3+1,\;s_1+s_2\geq s_3+1}
(and a certain $Z_2$ selection rule). Of course this can not
be the whole story, since \fusion\ is not symmetric
under permutations of $(1,2,3)$. Even if we symmetrize, it seems that
we have lost the truncation of the fusion rules \BPZ\ $r_1+r_2+r_3\leq
2p-1$, $s_1+s_2+s_3\leq 2p^\prime-1$. This is of course not the case; the
issue is the correct treatment of the flat space amplitudes.
The usual way one gets the three point couplings there is by factorization of
four point functions \DF. It is known that the direct evaluation
of the Feigin Fuchs integrals for the three point
function \threescr\  does not yield
the same results; rather one has to symmetrize, by
using the symmetry of flipping {\it any two} of the three vertices:
$V_{r,s}\rightarrow V_{p-r, p^\prime-s}$
\ref\DFL{V. Dotsenko and V. Fateev, Phys. Lett. {\bf 154B} (1985) 291.}.
This symmetry must be manifest in all $N$ point functions, and,
by construction, also after coupling to gravity. Thus we have to apply
this symmetry to \fusion. The result is \DFK:
\eqn\minimal{\langle\tilde T_{r_1,s_1}\tilde T_{r_2,s_2}\tilde T_{r_3, s_3}
\rangle=\mu^s
N_{(r_1,s_1), (r_2, s_2), (r_3, s_3)}}
where $N_{(r_i,s_i)}\in (0,1)$ are the flat space fusion numbers.
Eq. \minimal\ is compatible with matrix model results
\DIFK\ and generalizes
them considerably. Similar results for a subset of three
point functions were obtained in \ref\GL{M. Goulian and M. Li,
Phys. Rev. Lett. {\bf 66} (1991) 2051.}.

\noindent{\it 2.2.4. $N \geq 4$ point functions.}

In the previous subsections we have obtained the three point function
of the tachyon field. Remarkably, in two dimensional string theory
one can calculate {\it all} $N$ point functions.
The miraculous cancellations encountered above
will be seen here to be due to an interesting structure of the bulk
$N$ point functions.
For reasons to be explained
below we will restrict ourselves to $N$ point functions without screening
charges\foot{Although we believe the general case is not much harder.
It is also interesting, e.g. for the study of factorization
in 2D gravity coupled to minimal matter.}, where the conservation laws
take the form:
\eqn\consN{
\eqalign{ \sum_{i=1}^N k_i &= 2 \alpha_0 \cr
 s \alpha_+ + \sum_{i=1}^N \vert k_i - \alpha_0 \vert &= ({N \over 2} - 1)Q
 \cr}}
The correlator reads then:
\eqn\corN{
\eqalign{ A(k_1,..,k_N)&=(- \pi)^3 ({\mu \over \pi})^s \Gamma(-s)
\prod_{a=1}^s \int d^2w_a \prod_{i=4}^N \int d^2 z_i \cr
&\langle T_{k_1}(0) T_{k_2}(\infty) T_{k_3}(1)
\prod_{a=1}^s T_{0}(w_a) \prod_{i=4}^N
T_{k_i}(z_i) \rangle \cr}}
where we have fixed again the positions of three tachyons by $SL(2,{\bf C})$
invariance, and tuned
$k_i, \alpha_0$ such that $s$ is integer.
The free field correlator
is:
\eqn\ffN{
\eqalign{
\langle T_{k_1} T_{k_2} T_{k_3} \prod_{i=4}^N \int T_{k_i} \left[ \int
T_0 \right]^s \rangle &= \prod_{a=1}^s \int d^2 w_a \prod_{i=4}^N
d^2 z_i
\vert w_a \vert^{2 \delta_1} \vert 1-w_a \vert^{2 \delta_3}
\prod_{a<b} \vert w_a - w_b \vert^{4 \rho}  \cr
&\vert z_i \vert^{2 \theta_{1,i}}
\vert 1 - z_i \vert^{2 \theta_{3,i}} \prod_{i<j} \vert z_i - z_j \vert^{2
\theta_{i,j}} \prod_{i,a} \vert z_i - w_a \vert^{2 \delta_i} \cr}}
where:
\eqn\defin{\delta_i= -2 \alpha_+ \beta(k_i) \ \ \ ; \ \ \  \theta_{i,j}=k_i k_j
- \beta(k_i) \beta(k_j); \ \ \ \rho= - {\alpha_+^2 \over 2}}
Our experience from the previous cases suggests to study the
$(N-1, 1)$ kinematics\foot{All other $(n,m)$ kinematic regions
with $n,m>1$ give zero; this can be shown
by similar techniques to those used below.}:
\eqn\kin{ k_1,k_2,...,k_{N-1}> \alpha_0 \ \ \ ; \ \ \ k_N< \alpha_0<0 }
The conservation laws \consN\ lead to:
\eqn\kN{ k_N = {{N+s-3} \over 2} \alpha_+ + {\alpha_- \over 2} }
Anticipating the form of the result, we choose to
parametrize the momenta by the variables
$m_i={1 \over 2}(\beta(k_i)^2 - k_i^2)$, in terms of which:
\eqn\kinematics{\eqalign{
\delta_i &= \rho - m_i \ \ \ , \ \ \ i<N \cr
\delta_N &= -1- (N+s-3) \rho \cr
\theta_{i,j} &= -m_i - m_j \ \ \ , i<j<N \cr
\theta_{i,N} &= -1+(N+s-3)m_i \cr }}
Now \ffN\ does not look particularly simple. In fact, it is a special case
of the $N$ point amplitudes \Npoint,
which are certainly complicated. As we saw in section 2.1, the main
reason for the complications is the presence of poles \resN\
in all possible channels. We seem to have the same problem here:
upon observation, \ffN\ seems to have similar poles.
The main difference between \ffN\ and its higher dimensional analogues
is that in two dimensions {\it the residues of most of these poles
vanish!}
Consider for example the $(1,4)$ channel. The
first pole occurs when $\theta_{1,4}=-1$ (compare to \poles). The residue
of the pole (using \resN) involves the correlation function of
an intermediate tachyon at $k=k_1+k_4$ and $\beta=\beta_1+\beta_4$.
Plugging the on shell condition $\beta_i=k_i+\alpha_+$ into \defin\ we find
\eqn\intera{k={\alpha_-\over2}-\alpha_+;\;\;\beta={\alpha_-\over2}+\alpha_+}
Near the pole \ffN\ has the form (see \resN):
\eqn\residue{\langle T_{k_1}...T_{k_N}\rangle\simeq{1\over\theta_{1,4}+1}
\langle T_kT_{k_2}T_{k_3} T_{k_5}...T_{k_N}\rangle}
Now we proceed inductively. Suppose we have shown that all $M$ point functions
with $M\leq N-1$ satisfy
\eqn\polynom{\langle T_{k_1}...T_{k_M}\rangle=\prod_{i=1}^M\Delta(m_i)
P(k_1,..,k_M)}
with some polynomial in the momenta $P(k_i)$. We will soon show the same for
$M=N$, but in the meantime we can use \polynom\ to show that the residue
\residue\ vanishes:
$k$ \intera\ satisfies $m(k)=2$ and since $\Delta(2)=0$, plugging
\polynom\ in \residue\ we indeed get zero
for the residue of the pole at $\theta_{1,4}=-1$.
In other words, by two dimensional kinematics the on shell tachyon is
automatically at one of the (discrete) values of the momentum
for which the ``renormalization factor'' $\Delta(k)$ \redef\ vanishes.
Therefore, the residue \residue\ is zero.
This is of course markedly
different from the situation in higher dimensions.
One can argue similarly for the higher
poles at $\theta_{1,4}=-n$, $n\geq2$; for those we need a similar
property of the discrete oscillator states which we will derive in section 4.

The general poles were discussed in section 2.1. It is easy to
see that the residue \resN\ (with $\tilde k=2\alpha_0-\sum_ik_i$)
is almost always zero. For example, focussing on the
(first) poles that occur
when some of the $z_i\rightarrow0$ we have two classes of poles:

\noindent{}1) A subset of $z_i$, $i=4,..,N-1$ approach zero.
In this case, $T_{\Sigma_ik_i}$ in \resN\ has the property that
$m(\sum_ik_i)$ is a positive integer so that
$\Delta(m(\sum_ik_i))=0$ and the second term in the residue
\resN\ vanishes (using the induction hypothesis \polynom). The first term
is finite, thus the residue is zero.

\noindent{}2) A subset of $z_i$, $i=4,..,N-1$ and $z_N$ approach
zero. Here the second term in the residue \resN\ is finite
but the first term vanishes,
(again by
\polynom),
{\it except} when the subset
of $z_i, i=4,..,N-1$ is empty.
A similar structure occurs for the massive poles in all channels
(see section 4).

We see that the phenomenon underlying the vanishing of the residues
of the above poles is the special role of the states at
the discrete momenta ($\sqrt{2}
k\in Z$ for $c=1$). All the on shell intermediate states in \ffN\
occur at these momenta {\it in the wrong branch} $(\beta
<-{Q\over2})$. Their vanishing, advocated in \NATI\
is therefore crucial for the simplicity of the amplitudes
in 2D string theory. There seems to be a large symmetry
relating these states to each other which underlies this.

Thus the only poles with non vanishing residues in \ffN\ are those
coming from $z_N$ approaching one of the other vertices; since
$k_N$ is fixed \kN, this implies that although the interpretation
of the poles is the standard Veneziano one, they occur {\it only as a
function of individual external momenta} $k_i$
(or equivalently $m_i$) and not more complicated kinematic invariants.
I.e. the positions of poles in $m_1$, say, depends at most
on $s$ and not on the other $m_i$. The poles from $z_N\rightarrow\infty$
depend on $m_2$, which is a function of $m_1, m_3,.. m_{N-1}$
through the kinematic relation
\eqn\genkin{\sum_{i=1}^{N-1}m_i=1+\rho s}
One could ask, why aren't the residues of the poles in $m_i$ zero as well,
since as for all other poles, they can be seen to involve discrete
states in the wrong branch. The answer is that as we mentioned above,
the decoupling of these states is {\it only partial}. In the presence
of enough discrete states from the `right branch' it no longer occurs.
Indeed, the residue of the poles in $m_i$ involves three point functions
of the form $\langle V^{(-)}T_1T_2\rangle$ where $V^{(-)}$ is a discrete
state in the wrong branch (see section 4). The point is that
the tachyons $T_1,\; T_2$ are forced by kinematics
to be at one of the discrete
momenta in the `right branch', hence the residue is in
general non vanishing. We will return to the ``competition'' between
$V^{(+)}$ and $V^{(-)}$ in section 4.

Where are these poles located? Naively,
from \ffN\ first order poles in $m_1$ e.g. seem to
appear when $\theta_{1,N}=-l$ ($l=1,2,..$). However,
one can convince oneself that the residues always vanish except when $m_1=-n$
(and $-l=\theta_{1,N}=-1-(N+s-3)n$ \kinematics). This can be shown
either by noticing that the location of the poles is independent of $k_3,..,
k_{N-1}$, so we can take them to zero and use \resthree, or by showing
that only when $m_1=-n$, does the intermediate state describe an on shell
physical string state.
The residue of the corresponding pole
\resN\ is the correlation function
of $T_{k_2},.., T_{k_{N-1}}$ with one of the string states at level
$l-1=n(N+s-3)
$. A simple consistency check is that on shell discrete states\foot{
It is known
\ref\LZ{B. Lian and G. Zuckerman, Phys. Lett. {\bf 254B} (1990) 417.}
that in the minimal models $(c<1)$ there are no ``discrete oscillator
states'' in addition to the ``tachyon''
(although there are other new states \LZ). We find
such oscillator
states in intermediate channels here at generic $\alpha_0$
because we couple to gravity the Feigin Fuchs model \FF, \DF. The momentum
$k$ is continuous and there are discrete states as in the $c=1$
model. These states can be obtained by rotating the $c=1$
spectrum, as discussed below eq. \Srotated. A peculiar feature of these
states is that they do not have the form $V_me^{\beta\phi}$, but rather
depend non trivially on $\phi, X$. This is one of the indications
that these models should not be taken seriously, except as rotated $c=1$
\MPY.}
appear precisely when $l=-1-(N+s-3)n, n=1,2,3,4,...$ (see section 4).
Similarly, from the region $z_N \rightarrow\infty$
we find poles at $m_2=-n$ or expressing $m_2$ in terms of the other $m_i$
through \genkin,
the location of the poles is $m_1=1+\rho s-\sum_{i=3}^{N-1}m_i+n$.

Summarizing, if we consider $A(k_1,..,k_N)$ as a function of $k_1$, the pole
structure consists of first order poles at
\eqn\polesN{
\eqalign{
m_1&= -n \ \ \ ; \ \ \ n=0,1,2,3,..  \cr
m_1&= \rho s + 1 -\sum_{i=3}^{N-1} m_i +n \ \ \ ; \ \ \ n=0,1,2,3,.. \cr }}
In view of the result \polynom\ we are trying to prove, it is natural
to consider
\eqn\f{f(m_1,m_3,..m_{N-1})={\langle T_{k_1}...T_{k_N}\rangle\over
\prod_{i=1}^N\Delta(m_i)}}
We now know \polesN\ that all the poles of the numerator on the r.h.s. of \f\
are matched by similar poles of the denominator. Thus if $f$ \f\ is to
have any poles, they must come from zeroes of the denominator, which
are not matched by similar zeroes of the numerator. Of course, the denominator
has simple zeroes at $m_i=l$ ($l=1,2,3,..$). We will next show that the
correlator \corN\ also vanishes for these values of the momenta.
The simplest way to see that is to use \polynom\ recursively. We know that
$A(k_1,..k_N)$ has the form
$A=\prod\Gamma(m_i)g(k_1,..k_N)$.
We have to prove that $g=0$ whenever $m_i=l$. To do that
we can use the OPE of, say, $T_{k_{N-1}}$
and $T_{k_N}$ and focus on the residue of the pole at $m_{N-1}=-n$ which
is given by an $N-1$ point function, which vanishes for $m_1=l$ by \polynom.

{}From another point of view, by the standard DDK argument \DDK, the vanishing
of $T_{\alpha_0}$ (which is the first case $l=1$ where we want $A$ to vanish)
is
equivalent to KPZ scaling of correlation functions involving the operator
$\phi\exp(-{Q\over2}\phi+i\alpha_0 X)$. Thus it is good news that
$T_{\alpha_0}$ vanishes\foot{ In $D>2$ this is no longer the case:
the operator $\exp (ik\cdot X-{Q\over2}\phi)$
(with appropriate $k^2$) does not vanish in \szero; KPZ
scaling breaks down, and the structure of the theory is more complicated.}.
To show vanishing of $T^{(l)}\equiv T(m_k=l)$ for $l>1$ given vanishing of
$T^{(1)}=T_{\alpha_0}$, one can use the fact that $T^{(l)}=T_{\alpha_-\over2}
T^{(l-1)}$ recursively. We leave the details of this argument to the reader.

This concludes the proof of vanishing of $A(k_1, ...,k_N)$ whenever
$m_i=l(=1,2,3,..)$. Returning to \f, we have shown that $f$ is an entire
function of the $m_i$. To completely fix it we show that it is
bounded as $\vert m_i\vert\rightarrow\infty$. Consider, e.g. the $m_1$
dependence. For large $\vert m_1\vert$, \ffN\ is dominated by
$z_i, w_a\simeq1$. To blow up this region we redefine $z_i=\exp({x_i\over
m_1})$ and $w_a=\exp({y_a\over m_1})$ and estimate \ffN. We find
$f(m_1,..)\rightarrow {\rm const}$. Thus considered as a function of
$m_1$, $f$ is analytic everywhere and bounded at $\infty$.
Therefore,
it is independent of $m_1$. Repeating
the argument for the other $m_i$ (or by symmetry) $f$ is
independent of all $m_i$. It may only depend on $N, s$. But since
it is independent of $k_i$, we can set $k_3,..,k_{N-1}=0$
keeping $s,N$ fixed, and calculate $f(s,N)$ from \threept.
Plugging the result back in \f\ we finally find the $N$ point function
of the renormalized fields $\tilde T$ \redef:
\eqn\rfN{ \langle \tilde T_{k_1}...\tilde T_{k_N}
\rangle=(\partial_{\mu})^{N-3} \mu^{s+N-3} }
Notice that \rfN\ has the form \polynom\ as promised. As
explained above, for $s\not\in Z_+$ we still get
\rfN, assuming as before that the exact result
is a polynomial in momenta $k_i$.

Eq. \rfN\ completes the first two steps of the procedure described
in section 2.1.
Its region of validity is tied to the region of
convergence of \ffN. One can show that the latter converges
whenever $m_i>0$, $i=1,..,N-1$ (with the relation \genkin). In the
next subsection we will describe the correct continuation of \rfN\
to all momenta and find an interesting kinematic structure.

Finally, \rfN\ can be compared to matrix model results. We do that
in Appendix A and find agreement between the different
approaches.

\noindent{\it 2.2.5. The analytic structure of the $N$ point functions.}

So far we have treated the amplitudes in non critical string theory
using critical string techniques. We have found that the Shapiro -- Virasoro
amplitudes \Npoint, which are defined for arbitrary $D$, are actually
calculable for $D=2$ due to simplifications in the dynamics of the
theory. This involved two elements: we have used the fact that Liouville
momentum is conserved in the bulk, and continued the amplitudes
analytically from the region where they converge. In particular, in the
process we have ignored the requirement $\beta>-{Q\over2}$ ($E>0$)
discussed above.
We have shown that the amplitudes thus obtained have a standard
space time interpretation, although there are interesting symmetries
special to two dimensions, which make them simple. The set of bulk ($s=0$)
amplitudes defines therefore a consistent S -- matrix in the sense
of critical string theory. We have also begun in previous subsections
to extend the result to non integer $s$, obtaining \rfN. Since
no direct methods to evaluate such amplitudes are available, we had to
invoke a space time principle; the assumption that the $\tilde T$
amplitudes are described by a local two dimensional field theory,
and are polynomial in momenta. As it turns out, this
assumption rules out a naive analytic continuation of \rfN\ to all
momenta. Our next task is to understand the general structure
of the correlators by deriving constraints which Liouville
theory places on the form of this space time field theory
and propose the general correlators.

The basic property we will use is that when $s\not\in Z_+$ the Liouville
interaction is crucial, as explained in section 2.1, and momentum
is not conserved; we have:
\eqn\LOPE{\exp(\beta_1\phi)\exp(\beta_2\phi)=\int d\beta\exp(\beta\phi)
f(\beta,\beta_1,\beta_2)}
$f$ is an OPE coefficient. We did not specify the contour of integration
over $\beta$ in \LOPE\ since it is not known. In \NATI\ it has been argued
(based on space time considerations) that the amplitudes should be defined
by summing over macroscopic states $\beta=-{Q\over2}+ip$, $p\in R$.
We will adopt this procedure here.

Consider\foot{We thank N. Seiberg for discussions on this issue.}
the region of the moduli integrals in a generic tachyon
amplitude \tachcor\ where, say, $T_{k_1}(z)\rightarrow T_{k_2}(0)$.
The contribution of the region $z\rightarrow0$ to the amplitude
is given by \NATI:
\eqn\pl{\int_{\vert z\vert<\epsilon}d^2z\int_{-\infty}^\infty
dp(z\bar z)^{{1\over2}p^2+{1\over2}(k_1+k_2-\alpha_0)^2-1}f(p, \beta_1,
\beta_2)}
Assuming that we may interchange the order of integration
over $z,p$ we obtain\foot{More generally, if the matter sector OPE
produces an intermediate state of dimension $\Delta$,
we have: $\int {dpf(p,\beta_1,\beta_2)\over p^2+2(\Delta-{c-1\over24})}
\langle V_\Delta...\rangle$.
The only singularities occur at $E=\sqrt{2(\Delta-{c-1\over24})}\rightarrow
0$.}:
\eqn\pint{\int d^2z\langle T_{k_1}(z)T_{k_2}(0)...\rangle
\simeq\int_{-\infty}^\infty dp{f(p,\beta_1,\beta_2)\over p^2+(k_1+k_2
-\alpha_0)^2}\langle T_{k_1+k_2}...\rangle}
Now for fixed $p$, \pint\ has the familiar form from critical string
theory; we find a pole corresponding to the intermediate state
$T_{k_1+k_2}$. The fact that Liouville momentum is not conserved
and we have to sum over all $p$'s may turn this pole into a cut:
\pint\ depends on $\vert k_1+k_2-\alpha_0\vert$. Thus we expect
cuts whenever some of the momenta $\{k_i\}$ in \tachcor\ satisfy\foot{
Note incidentally that the integral representation always diverges
before any intermediate momentum gets to $\alpha_0$: if e.g.
$\sum_{i=1}^nk_i=\alpha_0$, $\sum_{i=1}^nm_i=1-\rho(n-1)$. Using \genkin\
we find that $\sum_{i=n+1}^{N-1}m_i=\rho(s+n-1)<0$. But the integral
representation converges iff all $m_i>0$. Thus the integral representation
is not useful to study the behaviour near the cuts.}
$\sum_ik_i\equiv p\rightarrow\alpha_0$. How can we make this more precise?
A major clue comes from comparing an amplitude with an insertion
of a puncture $P=T_{k=0}$ to the amplitude without it. By
KPZ scaling \newA, \Spar\ we have:
\eqn\PTTT{\langle PT_{k_1}...T_{k_N}\rangle=\left[-{\alpha_-\over2}
\sum_{i=1}^N\vert k_i-\alpha_0\vert-({N\over2}-1)(1+{\alpha_-^2\over2})
\right]\langle T_{k_1}...T_{k_N}\rangle}
Thinking of \PTTT\ as a relation between tree amplitudes in the purported
space time field theory reveals its essential features: we can insert
the puncture $T_{k=0}$ into the tree amplitude $\langle T_{k_1}...T_{k_N}
\rangle$ either by attaching it to one of the $N$ external legs, thus adding
an internal propagator of momentum $k_i+0=k_i$ or inside the diagram.
The first term (the sum) on the r.h.s. of \PTTT\ corresponds to the first
possibility; we can read off the propagator $-{\alpha_-\over2}\vert
k-\alpha_0\vert$. The second term corresponds to the second possibility,
and our remaining goal is to make it too more explicit.

The propagator we find is related to the two point function as usual; it
should be proportional to the inverse two point function (recall
that the correlation functions \tachcor\ have the external propagators
stripped). Indeed, by integrating \finalthree\ once (first putting
$k_2=0$), we find:
\eqn\twopt{\langle \tilde T_k\tilde T_{2\alpha_0-k}\rangle=-{1\over\alpha_-
\vert k-\alpha_0\vert}}
so that the propagator in \PTTT\ is ${1\over2}(\langle \tilde T_k
\tilde T_{2\alpha_0-k}\rangle)^{-1}$. Now that we understand the propagator,
the only remaining problem is the specification of the vertices
in the space time field theory. The three point vertex is 1 by \finalthree.
To find the higher irreducible vertices we have to use the
world sheet -- space time correspondence. Consider, for example,
the four point function
\eqn\fourpt{\tilde A(k_1,..,k_4)=\int d^2z\langle\tilde T_{k_1}(0)
\tilde T_{k_2}(\infty)\tilde T_{k_3}(1)\tilde T_{k_4}(z)\rangle}
To integrate out the massive string states we separate the $z$ integral
in \fourpt\ into two pieces. One is a sum of three contributions
of intermediate tachyons from the regions $z\rightarrow0,1,\infty$.
By \pl, \pint\ we expect to get $-{\alpha_-\over2}\vert k_1+k_i-\alpha_0
\vert$ from those. The rest of the $z$ integral is the contribution of massive
states; it gives a new irreducible four particle interaction
(which we will denote by $A_{1PI}^{(4)}$) for the
tachyons. The crucial observation that allows one to calculate this term
is that the contribution of the massive modes is analytic in external momenta.
This can be understood from several different points of view; from Liouville,
\pl, \pint\ imply that only intermediate states with $E\rightarrow0$
cause non analyticities of the amplitudes. But the massive states
occur only at discrete momenta (and energies) which are never close to zero.

This observation allows us to calculate $A^{(4)}_{1PI}$; we write
\eqn\fourpnt{A({k_1,..,k_4})=-{1\over2}\alpha_-
\left(\vert k_1+k_2-
\alpha_0\vert+\vert k_1+k_3-\alpha_0\vert+\vert k_1+k_4-\alpha_0
\vert\right)+A_{1PI}^{(4)}}
and now use the fact that we actually know $\tilde A(k_1,..,k_4)$
whenever, say, $k_1,k_2,k_3>0, k_4<\alpha_0$. In that
kinematic region we can compare the result \rfN\ with \fourpnt\ and find
\eqn\aonepi{A_{1PI}^{(4)}=-{1\over2}(1+{\alpha_-^2\over2})}
But now, for $A_{1PI}^{(4)}$ we know that
we can use analytic continuation
through the zero energy cuts, since by general arguments it must be
analytic in $k_i$. Of course this immediately implies that \aonepi\
is the correct irreducible four tachyon interaction everywhere. This
concludes the derivation of the tachyon four point function \fourpt.
A few comments about \fourpnt, \aonepi\ are in order:

\noindent{}1) The irreducible vertices for three and four tachyons
were found to be constant. This is {\it not} general. We will soon
see that for $N\geq5$ $A_{1PI}^{(N)}$ is a
highly non trivial (analytic)
function of the momenta.

\noindent{}2) For $c=1$ \fourpnt\ agrees with matrix model results
\MOORE, \INTER.

\noindent{}3) It is interesting to consider the cuts \fourpnt\ in the
case of the bulk amplitudes $s=0$ (since then the Liouville
momentum is conserved). For $d=c=1$ ($\alpha_0=0$) the only non zero
amplitudes are those with (e.g.) $k_1,k_2,k_3>0, k_4<0$. We can never
pass through $k_i+k_j=0$ because of kinematics. Therefore the cuts
\fourpnt\ are invisible in the bulk. This is no longer the case
for $c\not=1$. There we have $k_1,k_2,k_3>\alpha_0, k_4<\alpha_0<0$
and (e.g.) $k_1+k_2=\alpha_0$ is not on the boundary of this region.
What is the interpretation of the cuts then? We no longer have \LOPE--
the Liouville momentum is conserved in the bulk. However, as explained
above, the integral representation starts diverging before we get to
$k_1+k_2=\alpha_0$ (from $k_1+k_2>\alpha_0$). This is crucial for
consistency; we learn that when the integral representation diverges
we shouldn't use the naive continuation but rather use the space time
field theory as a guide, a point of view emphasized above.

\noindent{}4) We can now come back to the relation between the $\alpha_0
\not=0$ model and the two dimensional string mentioned below \Srotated.
We see \fourpnt\ that even for $s=0$ where naively the amplitudes in the
two cases are related by a rotation, this is not the case; the region
$k_i>\alpha_0$ is transformed to $\tilde k_i>0$ but the amplitudes \fourpnt\
do not transform accordingly. When the integral
representations diverge they are defined
in a different way in the two cases. However we see that both situations
are described by essentially the same two dimensional field theory in
space time.

\noindent{}5) Another curious feature of the $c<1$ ($\alpha_0\not=0$)
models is that the screening charges $V_{d_\pm}$ in \Srotated\ are
{\it not} treated on the same footing as the tachyon, despite
the fact that they are naively tachyon vertices of momenta $d_\pm$.
To see that one can compare the three point functions with screenings
to the $N$ point functions without screenings. For example, comparing
\rescr\ with $n=1$, $m=0$ to the four point function \fourpnt\
with one of the momenta equal to $d_+$ we find that in general the two differ.
Again, this is consistent, since the screening charges lie outside
of the region of convergence of the integral representation
\ffN, however the full implications of this observation are unclear to us.
These complications are also the reason why $N\geq4$
point functions with screening
are harder to obtain.

It is now clear how to proceed in the case of $N$ point functions. We assume
that we know already $A_{1PI}^{(4)}$,.., $A_{1PI}^{(N-1)}$.
Then we write
all possible tree graphs with $N$ external legs, propagator $
-{\alpha_-\over2}\vert k-\alpha_0\vert$ and vertices
$A_{1PI}^{(n)}$
$(n\leq N-1)$ and add an unknown new irreducible
vertex $A^{(N)}_{1PI}
(k_1,..,k_N)$. The interpretation in terms of integrating out massive
states is as before. $A^{(N)}_{1PI}$ is again analytic in $\{k_i\}$
and we can fix it by comparing the sum of exchange amplitudes (reducible
graphs) and $A_{1PI}^{(N)}$
to the full answer \rfN\ in the appropriate
kinematic region \kin. This fixes $A_{1PI}^{(N)}$
in the above kinematic
region. Then we use analyticity of
$A^{(N)}_{1PI}$ to fix it everywhere. The
outcome of this process is the determination
of the amplitudes in all kinematic regions given their values
in one kinematic region.

In principle, the procedure we have given above can be implemented
to find $A^{(N)}_{1PI}$,
in very much the same fashion as we have derived
$A^{(4)}_{1PI}$ above.
However, it is more convenient to use a different
technique, which we will describe next.

\noindent{\it 2.2.6. Irreducible $N$ point functions.}

We are faced with a kind of ``inverse problem'': given the set
of amplitudes $\langle\tilde T_{k_1}..\tilde T_{k_N}\rangle$
\rfN\ in the kinematic region\foot{We will restrict ourselves to the case
$c=1$ in this subsection.}
$k_1,..,k_{N-1}>0, k_N<0$, find the set of irreducible vertices which
together with the propagator $\vert k\vert\over\sqrt2$ give these amplitudes
in the appropriate kinematic region. It is important that the vertices are
analytic in $\{k_i\}$. It is very useful to Legendre transform: the
generating functional $G(j)$ for connected Green's functions
has the form
\eqn\cgf{ e^{- G(j)} = \int {\cal D} \psi e^{- S(\psi) + \int j \psi}}
where the action $S$ is given by
\eqn\actpsi{ S(\psi)=
-\sum_{n=2}^{\infty} {1 \over n!}\int dk_1..dk_n \psi(k_1)..\psi(k_n)
\delta(k_1+...+k_n)
A_{1PI}^{(n)}(k_1,..,k_n)}
and $A^{(2)}_{1PI}=
-{\sqrt{2}\over\vert k\vert}$. At tree level the function
$G(j)$ reads
\eqn\funcgreen{
G(j)= -\sum_{n=2}^{\infty} { 1 \over n!} \int dk_1..dk_n j(k_1)..j(k_n)
\delta(k_1+..+k_n) \langle \psi(k_1)...\psi(k_n) \rangle_c }
The connected correlators $\langle\psi(k_1)..\psi(k_n)\rangle_c$
are equal to $\tilde A(k_1,..,k_n)=\langle\tilde T_{k_1}..\tilde T_{k_n}
\rangle$ up to insertion of external propagators, which appear in the
former and are stripped off in the latter. It is natural to redefine
$j(k)\rightarrow {\sqrt{2}j(k)\over\vert k\vert}$ on
the r.h.s. of \funcgreen\ after which
\eqn\greenfunc{
G(j)=
- \sum_{n=2}^{\infty} {1 \over n!} \int dk_1..dk_n
j(k_1)..j(k_n) \delta(k_1+..+k_n) {\tilde A}(k_1,..,k_n) }
It is well known that the saddle approximation in \cgf\ is exact
at tree level; therefore
\eqn\legendre{ -G(j)=-S(\psi) + \int {\sqrt{2}\over
\vert k\vert}j \psi \vert_{{\sqrt{2}\over\vert k\vert}j=S'(\psi)} }
By duality of the Legendre transform we also have
\eqn\dualeg{ -S(\psi)=-G(j) - \int
{\sqrt{2}\over\vert k\vert}j \psi \vert_{{\sqrt{2}\over\vert
k\vert}\psi=-G'(j)}\ , }
which implies that $S(\psi)$ is the generating functional for connected
tree level Green's functions arising from the action \greenfunc.
In other words, the irreducible amplitudes $A^{(n)}_{1PI}$
in \actpsi\ play now the role of {\it amplitudes}
generated by Feynman rules with propagator
of opposite sign $-{\vert k\vert\over\sqrt2}$
and the full amplitudes $\tilde A(k_1,..,k_N)$ playing the role
of vertices. To use our knowledge of $\tilde A$ in the kinematic
region $k_1,..,k_{N-1}>0$ we can now calculate these ``dual'' amplitudes
in that region of momentum space. Of course, we must first verify that
if the external momenta lie in this kinematic region, then for all
internal vertices in all possible
Feynman diagrams there are precisely $n-1$ positive and 1 negative
incoming momenta (since otherwise the ``dual vertices'' are unknown).
One can easily convince oneself that this is the case. After calculating
$A^{(N)}_{1PI}$ from these Feynman rules, we can continue them analytically
to all $k$ using their analyticity.

The problem of calculating
$A_{1PI}^{(N)}$ has been reduced to evaluation
of tree level Feynman diagrams. The general expressions are complicated;
we discuss them in Appendix B. Here we will
illustrate the kind of results one gets by giving two typical examples:
\eqn\threenz{ A_{1PI}^{(N)}(k_1,k_2,k_3,k_4=0,..,k_N=0)=
(\partial_{\mu})^{N-3}
\left\{{1\over\mu}
\prod_{i=1}^{3} { 1 \over \cosh( {k_i \over \sqrt{2}} \log \mu)}\right\}
\bigg\vert_{\mu=1} }
\eqn\fournz{
\eqalign{
A_{1PI}^{(N)}(k_1,k_2,k_3,k_4,0,..,0)&=\cr
(\partial_{\mu})^{N-4}
\mu^{-2} \left( \prod_{i=1}^4 {1 \over {\cosh({k_i \over \sqrt{2}}\log \mu)}}
\right)
&\left( -1 -\mu \partial_{\mu} \log \prod_{1 \leq i < j \leq 3}
\cosh({{k_i+k_j} \over \sqrt{2}} \log \mu) \right)\bigg
\vert_{\mu=1} \cr}}
Notice that as expected, the irreducible amplitudes \threenz,
\fournz\ are analytic in $\{k_i\}$. A general proof of this statement
is given in Appendix B.
The discussion in this subsection is closely related to the
structure at $k=0$ discussed in
\ref\BKS{E. Brezin, V. Kazakov and N. Seiberg, unpublished as quoted
in V. Kazakov, preprint LPTENS 90/30 (1990).}.

\bigskip

\newsec{Two Dimensional Fermionic String Theory.}

\subsec{The model.}

We will not repeat the general considerations
of section 2.1 for the fermionic case as they are quite similar.
Instead, we will turn directly to the situation in two dimensions
which is the case of interest to us here.

The matter system is in this case one superfield
\eqn\supermat{X=x+\theta\psi_x+\bar\theta\bar\psi_x+i\theta
\bar\theta F_x}
which we want to couple to supergravity. As in the bosonic case
it is convenient to generalize by turning on a background
charge for $x$, which is therefore governed by the action \ma.
The fermions $\bar\psi_x\;(\psi_x)$ are free, left (right) moving.
Similarly, we have a Liouville superfield
\eqn\superliouv{\Phi=\phi+\theta\psi+\bar\theta\bar\psi+i\theta
\bar\theta F_l}
related to the conformal factor of the metric and the gravitino
field in superconformal gauge. $\Phi$ is governed by the action
\POL:
\eqn\SLaction{S_{SL}={1\over2\pi}\int d^2z\int d^2\theta\left[
D\Phi\bar D\Phi+
2\mu\exp(\alpha_+\Phi)\right]}
where $D=\partial_\theta+\theta\partial_z$, and we have dropped
curvature couplings \SL.
The central charge of $X$ \supermat\ is $\hat c={2\over3}c=1-
8\alpha_0^2$ and that of $\Phi$, ${\hat c}_{SL}=1+2Q^2$, where
\ref\DHK{J. Distler, Z. Hlousek and H. Kawai, Int. Jour.
Mod. Phys. {\bf A5} (1990) 391.}:
\eqn\superQ{Q=\sqrt{9-\hat c\over2};\;\alpha_+=-{Q\over2}+
\vert\alpha_0\vert}
Since we are making a non chiral GSO projection, we have two sectors
in the theory: (NS, NS) and (R, R) \GSW.
The (NS, NS) sector contains one field theoretic degree of freedom,
the ``tachyon'' center of mass of the string, whose vertex operator
is given by
\eqn\supertach{T_k=\int d^2\theta \exp(ikX+\beta\Phi);\;
\beta+{Q\over2}=|k-\alpha_0|}

Bulk correlation functions of \supertach\ are calculated by integrating
over the locations of $N-3$ of the $T_k$, and inserting two of the
remaining vertices as lower components:
\eqn\supercor{\eqalign{
A(k_1...,k_N)=&(-\pi)^3 \int d^2\theta_1\prod_{j=4}^N
\int d^2z_j\int d^2\theta_j
\langle
\exp(ik_1X(0)+\beta_1\Phi(0))\cr
&\exp(ik_2x(\infty)+\beta_2\phi(\infty))
\exp(ik_3x(1)+\beta_3\phi(1))
\exp(ik_jX(z_j)+\beta\Phi(z_j))\rangle\cr}}
As before, the cosmological term in the action
\SLaction\ is the zero momentum tachyon. This presents the following
subtlety. We can write $T_k$ in components as:
\eqn\comp{T_k=\exp(ikx+\beta\phi)\left[(ik\psi_x+\beta\psi)(ik\bar
\psi_x+\beta\bar\psi)+i\beta F-kF_x\right]}
The auxiliary fields $F_x, F$ have delta function propagators
(in the free theory \SLaction); this can cause divergences
of the form $\delta^2(z)|z|^a$
in the OPE of the fields $T_k$ \comp. This is a familiar issue
in fermionic string theory
\ref\GRSE{M. Green and N. Seiberg, Nucl. Phys. {\bf B299}
(1988) 559; M. Dine and N. Seiberg, Nucl. Phys. {\bf B301}
(1988) 357.}; we have two possible ways to proceed:

\noindent{}1) Calculate everything at generic momenta. In this case
we can set the auxhiliary fields $F=0$, since we can continue
analytically from a region in momentum space where the contact
terms do not contribute.

\noindent{}2) If we must calculate at some given momentum, we have
to carefully regulate the divergences in a way compatible with
world sheet supersymmetry (SUSY).
In particular we must keep $F$ \GRSE.

The second procedure is in general difficult to implement, especially
in the presence of Ramond fields. Therefore, we will use the first one. Note
that in this case we will not be able to perform the generalization
of \ampiii\ here\foot{Indeed, we are not aware of the existence of
(analogous) calculations for the Feigin Fuchs representations
of the supersymmetric minimal models
\ref\FR{D. Friedan, Z. Qiu and S. Shenker,  Phys. Lett. {\bf 151B}
(1985) 37; M. Bershadsky, V. Knizhnik and M. Teitelman, Phys.
Lett. {\bf 151B} (1985) 131.}.}.

The Ramond (R, R) sector gives rise to another (massless) field theoretic
degree of freedom, whose vertex operator can be constructed
using \ref\FMS{D. Friedan, E. Martinec and S. Shenker, Nucl. Phys.
{\bf B271} (1986) 93.}. First, we bosonize the fermions $\psi_x,\psi$
as:
\eqn\boson{\psi={1\over\sqrt2}(e^{ih}+e^{-ih});\;
\psi_x={1\over\sqrt2}(ie^{ih}-ie^{-ih})}
where $\langle h(z)h(w)\rangle=-\log(z-w)$, and similar expressions hold
for the left movers (which we will suppress below).
The R vertex is given by
\eqn\Vminus{V_{-{1\over2}}=\exp\left(-{1\over2}\sigma+{i\over2}
\epsilon h+ikx+\beta\phi\right);\;\beta=-{Q\over2}+\vert k-\alpha_0\vert}
$V_{-{1\over2}}$ is the fermion vertex in the ``$-{1\over2}$ picture''.
There is an infinite number of versions of $V$ in different pictures
(see \FMS).
$\sigma$ in \Vminus\ is the bosonized ghost current
and $\epsilon=\pm1$. The mass shell condition
for $\beta$ in \Vminus\ does not ensure BRST invariance in this
case. Imposing invariance w.r.t. the susy BRST charge,
$Q_{\rm susy}=\oint\gamma T_F$ with $T_F=\psi_x\partial x+\psi\partial\phi
+Q\partial\psi-2i\alpha_0\partial\psi_x$, we find
\eqn\dirac{\beta+{Q\over2}=-\epsilon(k-\alpha_0)}
This is the two dimensional Dirac equation in space time.
Correlation functions
involving Ramond fields are constructed using standard rules \FMS.
Defining $T_k^{(-1)}=\exp(-\sigma+ikx+\beta\phi)$, correlation functions
with two (R, R) fields have the general form
\eqn\twoV{A_{2V}(k_1,...,k_N)=\langle
V_{-{1\over2}}V_{-{1\over2}}T_{k_3}^{(-1)}T_{k_4}..T_{k_N}\rangle}
those with four (R, R) fields\foot{Only correlators with an even
number of Ramond fields can be non zero due to a $Z_2$ symmetry.}
\eqn\fourV{A_{4V}(k_1,...,k_N)=\langle V_{-{1\over2}}V_{-{1\over2}}
V_{-{1\over2}}V_{-{1\over2}}
T_{k_4}..T_{k_N}\rangle}
where $N-3$ of the vertices are always
integrated. For correlators with more than four
(R, R) fields we need $V_{+{1\over2}}$; we will not consider
those here, but give its form for completeness:
\eqn\Vplus{V_{1\over2}=(2\epsilon k+Q)\exp({\sigma\over2}
+{3\epsilon\over2}ih+ikx+\beta\phi)+
(\partial\phi-i\epsilon\partial x+2\alpha_0-\epsilon Q)
\exp({\sigma\over2}-{\epsilon
\over2}ih+ikx+\beta\phi)}

\subsec{The massless S -- Matrix.}

Most of the features of the discussion of the wave function \ps, the
$\phi$ zero mode integration \newA\ and its space time interpretation,
can be borrowed for the supersymmetric case. The only modification
of \newA\ needed is replacing bosonic correlators by fermionic ones
(replacing fields by superfields \supermat, \superliouv, moduli
by supermoduli, etc) as well as adding the new (R, R) field
$V$. Since, as explained above, we are forced to analytically continue
in momenta in order to ignore contact terms, we concentrate below
on the case $s=0$ in \newA\ (which is in any case the most general bulk
amplitude). In the next two subsections we first consider
the S -- matrix of the tachyon $T$ and, then that of the Ramond
field $V$.

\noindent{\it 3.2.1. Tachyon scattering in fermionic 2D string theory.}

It is useful to start with \supercor\ for the case $N=4$ (and $s=0$);
putting $F=F_x=0$ in \comp\ we find:
\eqn\superszero{A_{s=0}(k_1,..,k_4)=\pi^4 \prod_{i=1}^3
{\Gamma(k_4\cdot k_i-\beta_4
\beta_i+1)\over\Gamma(\beta_4\beta_i-k_4\cdot k_i)}}
This formula, which is superficially identical to \szero, is of course
true (as there) for all values of the dimension of space time. The poles
reflect again the presence of massive string states, which in two dimensions
are restricted to special momenta ($k\in Z$). To study the simplifications
in $D=2$, we use \supertach\ and find that:

\noindent{}1) In the ``(2,2)'' kinematics $k_1,k_2>\alpha_0$,
$k_3,k_4<\alpha_0$, the amplitude \superszero\ vanishes.
This seems peculiar, since we expect poles with finite
residues in the $s,t,u$ channels (as in \resN).
However, the poles in the (say) $u$ channel are absent because
the intermediate momentum is fixed by kinematics, while those in the
$s,t$ channel cancel among themselves (precisely as in the bosonic case).

\noindent{}2) For (3,1) kinematics, $k_1,k_2,k_3>\alpha_0, k_4<\alpha_0$
(or vice versa):
$A_{s=0}(k_1,..,k_4)=\prod_{i=1}^3\Delta(\tm_i)$,
where
\eqn\tmi{\tm_i={1\over2}\beta_i^2-{1\over2}k_i^2+{1\over2}}
In fact, since in this case kinematics forces $\tm_4=0$, we can, as
in \resthree, absorb the $\log\mu$ into an infinite
factor in the amplitude and write:
\eqn\sfour{A_{s=0}(k_1,..,k_4)=\prod_{i=1}^4(-\pi) \Delta(\tm_i)}
Now, eq. \sfour\ is equivalent to \superszero\ in all kinematic regions
(recall that a finite $A$ \sfour\ is interpreted as zero in the bulk
-- we need a pole to produce the $\log\mu$ implicit in \superszero).
The form of \sfour\ is suggestive (compare e.g. to \resthree).
We recognize many of the familiar features from the bosonic case;
e.g. the first zero at $\tm_i=1$ occurs at $\beta=-{Q\over2}$
(zero energy) and has a similar interpretation. The poles at
$\tm_i=0,-1,-2,..$ occur (for $\hat c=1$) at $|k|=1,2,3,...$, which is again
the set of momenta where oscillator states exist
(see section 4). Our next goal is
to show that the simple structure of \sfour\ persists for higher
point functions.

Thus we return to the $N$ point function \supercor\ with $s=0$. It is clear
from the discussion of the four point function above that the interesting
kinematics to consider is $(N-1, 1)$ (the rest will vanish
identically). We choose it to be the same as in the bosonic case \kin;
other regions can be treated similarly. Energy/momentum conservation
leads to $k_N={N-3\over2}\alpha_++{1\over2}\alpha_-$ (here we defined
$\alpha_-\equiv{1\over\alpha_+}$), or by \tmi, $\tm_N=-{1\over2}(N-4)$.
We expect to get the bulk divergence from an infinity of
$\Gamma(\tm_N)$, which happens only for even $N$. This is consistent
with \supercor: due to the (global) $Z_2$ R -- symmetry
$\psi\rightarrow-\psi, \bar\psi\rightarrow\bar\psi$,
\supercor\ is indeed zero
identically\foot{To avoid misunderstanding, we emphasize
that this does not necessarily mean that correlators of an odd number
of tachyons vanish, but only that they vanish {\it in the bulk}.}
for odd $N$. Therefore, we replace $N\rightarrow 2N$ in \supercor\ and
proceed. We have constructed the arguments in section 2 in such a way
that the generalizations are trivial. First one has to show that
the residues of most of the apparent poles in \supercor\ as groups
of $\{z_i\}$ get close, vanish. These residues have to do as before \resN\
with correlators involving physical states at the discrete
momenta $k\in Z$ and in the wrong branch.
Therefore we have to show decoupling of such states; this works precisely
as in the bosonic case (see section 4).
Assuming that, we have again only poles coming from
$z_{2N}$ approaching other $z_i$. Their locations are easily verified
to be $\tm_i=-l$ ($l=0,-1,...$) corresponding to intermediate states
of mass $m^2=(2l+1)(2N-3)$; only odd masses appear due to the $Z_2$
R -- symmetry mentioned above
($\psi\rightarrow-\psi$) under which the tachyon and all other
states with even $m^2$ are {\it odd}.

We define, in analogy with \f,
\eqn\superf{\tilde f(\tm_1,\tm_3,..,\tm_{2N-1})={A(k_1,..,k_{2N})\over
\prod_{i=1}^{2N}\Delta(\tm_i)}}
All the poles of the numerator $A$ are matched by poles of the denominator;
it is again necessary to show that $A$ vanishes whenever (say)
$\tm_1=1,2,3,...$ . This is the case for \sfour, and we can proceed
recursively as in the bosonic case, or use a symmetry argument relating
vanishing of $T(\tm_1=l)$ to that of $T(\tm_1=l+1)$ (see discussion
after \f). Therefore, $\tilde f$ \superf\ is an entire function
of $\tm_i$.
One can also show in complete parallel with the bosonic case that $\tilde f$
is bounded as $\vert \tm_1\vert\rightarrow\infty$ (say).
To do this we redefine
$z_i=e^{\xi_i\over m_1}$ in \supercor\ and (after some algebra)
find that $\tilde f\rightarrow{\rm const}$ as $\tm_1\rightarrow\infty$.
Since an entire function which is bounded at infinity is constant,
we conclude that $\tilde f$ depends at most on $N$ and the
central charge.

This concludes the evaluation of the bulk tachyon amplitudes;
the final result is \superf; $A(k_1,..,k_{2N})$ is proportional
to a product of ``leg factors'' up to a function $\tilde f$ of $N$,
$\hat c$.
In the bosonic case we could fix the function $f$ \f, the analog of $\tilde f$,
by using \resthree. This is not available to us here, but we can still
determine $\tilde f$ by a space time argument analogous to the one made in
the bosonic case.

The point is that regardless of whether we know $\tilde f(N)$ or not,
we have to perform now steps 2,3 of the general program of section 2.1.
We again make the assumption (which
is plausible, but was not derived neither in the bosonic case nor here) that
the massless amplitudes are governed by a 2D field theory (which now has
two fields), and furthermore that correlators in this theory are algebraic
in momenta. Eq. \superf\ (with $\tilde f=\tilde f(N, \hat c)$) is a highly
non trivial check of this idea. Using the above assumption,
we can find $\tilde f$ by calculating the two point function $
\langle T_kT_{2\alpha_0-k}\rangle$ for all $k$. The two point
function is (up to an unimportant constant) the inverse propagator, which we
can obtain by using KPZ scaling as in \PTTT. Repeating the same argument
here we find the propagator $-\alpha_-\vert k-\alpha_0\vert$. Thus the
two point function (in a convenient normalization) is
$\langle T_kT_{2\alpha_0-k}\rangle=-{1\over2\alpha_-\vert k-\alpha_0\vert}$.
This translates in \superf\ to
\eqn\tildef{\tilde f=(-\pi)^{2N}(2N-3)!}
The constant
can be determined by comparing to \sfour. It would be nice
to verify this result directly by computing $\tilde f(N)$ from the
integrals \supercor\ (for $N=2$ we have checked this form above \sfour).

As in the bosonic case, we can now obtain the general $N$ point functions
(any $s$). In fact, redefining
\eqn\sredef{\tilde T_k={T_k\over(-\pi)\Delta({1\over2}\beta(k)^2-{1\over2}
k^2+{1\over2})}}
we find that $\tilde T_k$ scattering is described by {\it the same
S -- matrix} as that of the bosonic tachyon \redef.
Some examples:
\eqn\NScor{\eqalign{
\langle\tilde T_{k_1}\tilde T_{k_2}\tilde T_{k_3}\rangle=&1\cr
\langle\tilde T_{k_1}\tilde T_{k_2}\tilde T_{k_3}
\tilde T_{k_4}\rangle=&-\alpha_-(|k_1+k_2-\alpha_0|+
k_1+k_3-\alpha_0|+|k_1+k_4-\alpha_0|)-{1\over2}(1+\alpha_-^2)\cr}}
etc. The cuts at $k_i+k_j=\alpha_0$ correspond to intermediate
tachyons\foot{Note that in the interacting theory, the symmetry
$\psi\rightarrow-\psi$ is broken by the interaction in \SLaction;
therefore, although the tachyon is odd under this symmetry, we
do have a non zero tachyon three point function, tachyon intermediate states
in the four point functions, etc.},
as in the bosonic theory. Eq. \NScor\
coincides with \fourpnt, \aonepi\
after making the identification
$k_{\rm fermionic}={1\over\sqrt2}k_{\rm bosonic}$,
$(\alpha_-,\alpha_+,\alpha_0)_{\rm fermionic}=
{1\over\sqrt2}(\alpha_-,\alpha_+,\alpha_0)_{\rm bosonic}$.
The only difference is in the
external leg factors \redef, \sredef\ reflecting a different spectrum
of oscillator states. This is reminiscent of earlier ideas
\ref\DIDK{P. Di Francesco, J. Distler and D. Kutasov, Mod. Phys.
Lett. {\bf A5} (1990) 2135.} relating bosonic and fermionic
strings in two dimensional space time (although clearly
one needs much more information for a complete comparison of the two
theories). In the next subsection, we will study one aspect
of the fermionic theory which certainly has no counterpart in the bosonic
one: the dynamics of the Ramond field $V$.

\noindent{\it 3.2.2. Scattering of the Ramond field .}

We follow again the same steps as for the tachyon field $T_k$. First we
consider four point functions. In order to have a non zero
bulk four point function of two R fields and two tachyons, we must
choose both R particles to move in the same direction, say
to the right $k>\alpha_0$. Then the amplitude \twoV\
can be evaluated to give:
\eqn\VVTT{A=\pi^4 (\beta_4^2-k_4^2){\Gamma(k_1k_4-\beta_1\beta_4+{1\over2})
\Gamma(k_2k_4-\beta_2\beta_4+{1\over2})\Gamma(k_3k_4-\beta_3\beta_4+1)\over
\Gamma(\beta_1\beta_4-k_1k_4+{1\over2})
\Gamma(\beta_2\beta_4-k_2k_4+{1\over2})\Gamma(\beta_3\beta_4-
k_3k_4)}}
If both tachyons move left $k_3, k_4<\alpha_0$ ((2,2) kinematics),
\VVTT\ vanishes, while if the signature is (3,1) we find {\it again} \sfour\
with one modification; $\tm_i$ has the form \tmi\ for the NS particles
($i=3,4$) while for the Ramond field $V$:
\eqn\Rmi{\tm_i={1\over2}(\beta_i^2-k_i^2)}
In complete parallel with the previous cases, \sfour\ can now be verified
to describe {\it all} bulk four point functions involving an arbitrary
combination of R and NS fields (provided the correct $\tm_i$ \tmi, \Rmi\
are used). One has to remember that in fermionic string theory in addition
to the trivial momentum conservation $\delta(\sum_ik_i-2\alpha_0)$,
which is implied in all amplitudes, we also have a $Z_2$ selection rule:
a Kroenecker $\delta$ of the number of R fields
modulu two: only correlation
functions with an even number of $V$'s can be non zero. A non trivial
check of \sfour\ is the four R field scattering: according to \sfour\
we should get zero identically in the bulk. This can be verified directly
by computing the integrals.

The form \sfour, \Rmi\ of Ramond scattering has the following
interesting feature: the zero energy $(k=\alpha_0)$ states
\Vminus\ {\it do not} decouple, unlike the case of the tachyon \tmi,
despite the fact that their wave function \ps\ is not peaked at $\phi
\rightarrow\infty$. We saw in section 2 (see discussion following
eq. \f) that one way to understand
the decoupling of the zero energy
tachyon is KPZ scaling. At $k=\alpha_0$ there is an additional BRST
invariant tachyon state
$\phi\exp(-{Q\over2}\phi+i\alpha_0 X)$; KPZ
scaling of its correlation functions is equivalent to vanishing
of the operator
$\exp(-{Q\over2}\phi+i\alpha_0 X)$. That argument goes through in the
supersymmetric case: the operator
$\phi\exp(-\sigma-{Q\over2}\phi+i\alpha_0 x)$ is BRST invariant,
therefore $T_{\alpha_0}$ \supertach\ must decouple.
In the Ramond sector on the other hand,
the operator with an insertion
of $\phi$ at $\beta=-{Q\over2}$
{\it is not} BRST invariant,
as is easy to verify. Therefore,
$V_{k=\alpha_0}$ need not (and does not) vanish.
One can also understand the difference between
the situation between the NS and R sectors from a different point of
view\foot{We thank N. Seiberg for this argument.}.
The exact wave functions of the various states satisfy the WdW equation
\MSS. In the NS sector, the form of this equation is such that
if as $\phi\rightarrow\infty$, $\Psi(\phi)\rightarrow {\rm const}$,
then in the IR, ($\phi\rightarrow-\infty$), $\Psi(\phi)$
blows up. This means that the operator $\exp(-\sigma-{Q\over2}\phi+
i\alpha_0x)$ behaves like the operators with $E<0$ ($\beta<-
{Q\over2}$, see section 2) and should decouple.
In the Ramond sector, the form of the WdW equation allows
a zero energy
solution which is constant at large $\phi$, decays at $\phi
\rightarrow-\infty$,
and is normalizable. Thus in this case the zero energy state
behaves like the macroscopic states \NATI\ and need not decouple.

We now turn to $N$ point functions \twoV, \fourV.
All the steps are as in the previous two cases, so we will be brief.
The main issue is the analysis of poles and zeroes. This is performed
precisely as before: the residues of most of the poles vanish by using
\sfour\ recursively (as well as properties of the discrete states). The only
poles occur at $\tm_i\in Z_-$ (with the notation \tmi\ (NS), \Rmi (R))
and correspond to on shell intermediate states.
The zeroes are also treated as before; we leave the details to the reader.
We find again that $\tilde f$ \superf\ is an analytic function
of momenta ($\tm_i$); in a by now standard fashion we also show that
it is bounded as $\vert \tm_i\vert\rightarrow\infty$, hence it is independent
of the $\{\tm_i\}$. To determine $\tilde f$ we use space time arguments, as
for the tachyon. KPZ scaling \Spar\ allows us to read off the propagator
for the Ramond field, $-\alpha_-|k-\alpha_0|$, and consequently
the two point function $\langle V_kV_{2\alpha_0-k}\rangle=-{1\over
2\alpha_-|k-\alpha_0|}$. This fixes $\tilde f$ to be the same as before
\tildef.

We now have all the correlation functions involving Ramond fields
(we actually checked those involving up to four R fields, but showed
how to obtain all of them, and conjecture that the results are going to
agree as well). For example, after absorbing the external leg
factors as in \sredef\ (and for the R field as well), we have:
\eqn\Rcor{\eqalign{
\langle V_{k_1}V_{k_2}T_{k_3}\rangle=&1\cr
\langle V_{k_1}V_{k_2}T_{k_3}T_{k_4}\rangle=&-\alpha_-\left(|k_1
+k_2-\alpha_0|+|k_1+k_3-\alpha_0|+|k_1+k_4-\alpha_0|\right)
-{1\over2}(1+\alpha_-^2)\cr
\langle V_{k_1}V_{k_2}V_{k_3}V_{k_4}\rangle=&
\langle V_{k_1}V_{k_2}T_{k_3}T_{k_4}\rangle\cr }}
The space time interpretation is as before. The cuts correspond
to massless intermediate states
(with $VV\rightarrow T$, $VT\rightarrow V$, $TT\rightarrow T$),
and the contact terms to a new irreducible
interaction.

\subsec{Chiral GSO projection.}

In fermionic 2D string theory we have the option to make a chiral GSO
projection \KUS. For $D>2$ this is useful to construct stable (tachyon free)
string theories with space time fermions. In $D=2$ there are no tachyons,
but one may still make the projection. This is useful as a toy model
for higher dimensional (non) critical superstrings. We will briefly review
the construction of \KUS\ in $D=2$ and discuss some of the emerging
properties\foot{This subsection is based on
\ref\ND{D. Kutasov, G. Moore and N. Seiberg, unpublished.}
(see also
\ref\me{D. Kutasov, Princeton preprint PUPT-1277 (1991).}).
We will put $\alpha_0
=0$ for simplicity.}.

We start with the observation \KUS\ that the 2D fermionic string system,
which consists of two superfields \supermat, \superliouv\ has
a natural global $N=2$ superconformal symmetry. The $U(1)$ generator, which
connects the two supercurrents is $J(z)=i\partial h+2i\partial x$
(the cosmological term in \SLaction\ breaks this symmetry). There is a well
known procedure in the critical string implementing the GSO projection
in the presence of such a symmetry
\ref\FKSW{D. Friedan, A. Kent, S. Shenker and E. Witten, unpublished;
T. Banks, L. Dixon, D. Friedan and E. Martinec, Nucl. Phys.
{\bf B299} (1988) 613.}, which we
imitate here.
We define
\eqn\IZ{I(z)=\exp(-{1\over2}\sigma(z)-{i\over2}h(z)+ix(z))}
$I(z)$ is a holomorphic operator ($\bar\partial I=0$). Note
that it is BRST invariant \dirac.
We now project out all operators \supertach, \Vminus\ etc, which do not
have a local OPE with $I(z)$ \IZ. This removes some states from the
existing (NS, NS), (R, R) Hilbert spaces. By acting on the remaining
states with $I(z)$ we generate two new sectors, (R, NS) and (NS, R),
which contain space time fermions. Geometrically, the chiral GSO projection
corresponds to enlarging the gauge group on the world sheet
by a certain $Z_2$ R -- symmetry \ND.
The operator
\eqn\QS{Q=\oint I(z); \; Q^2=0}
generates target SUSY. Due to the low dimension (and lack of
time translation invariance) the SUSY generator $Q$ is a kind of BRST
operator (in higher dimensions one would find a ``space SUSY" algebra
in the transverse directions \KUS). How does the spectrum look after
the projection in $D=2$? It is convenient to analyze it chirally:

\noindent{}$\underline{\rm NS\;\; sector:}$
Requiring locality of the `tachyon' \supertach\ with $I(z)$ \IZ, we
find that only $T_k$ with $k\in Z+{1\over2}$ survive. In addition
we have all the discrete states with odd $m^2$, starting with $\partial
x$.

\noindent{}$\underline{\rm R\;\; sector:}$
Imposing locality of \Vminus\ with \IZ\ we find two solutions:
a) $\epsilon=-1,\; k\in Z_+$,
b) $\epsilon=+1,\; 0>k\in Z+{1\over2}$.

The cosmological constant operator $T_{k=0}$ \SLaction\ has been
projected out of the spectrum; it is very natural \KUS\ to set the
scale with $T_{1\over2}$, which preserves the $N=2$ symmetry. If we
add it to the action with coefficient $\mu$, all the operators
left in the theory have the interesting property that their correlation
functions scale as integer powers of $\mu$. This is very reminiscent
of the topological theory of $c=-2$ matter coupled to ordinary (bosonic)
gravity
\ref\TOP{
E. Witten, Nucl. Phys. {\bf B340} (1990) 281;
J. Distler, Nucl. Phys. {\bf B342} (1990) 523.}.
Superficially there are problems with a topological
interpretation of our theory: by using \superf, \Rmi\ for the Ramond
correlators we see that for half of the R states (those with
$\epsilon=+1$), most correlation functions blow up.
Also, the fact that only integer powers of $\mu$ appear in correlation
functions is spoiled by addition of $\partial x\bar\partial x$ to the action
(the scaling dimensions change continuously with the radius).
Despite these problems, there probably is a topological theory here.
The point is that we have not used the BRST like properties
of the operator $Q$ \QS. In \TOP, the topological theory
had in addition to the usual string BRST another gauged topological
symmetry. Perhaps we should add $Q$ to our $N=1$ superconformal
BRST charge. Doing that, requiring that
\eqn\QBRST{Q|{\rm phys}\rangle=0}
we find that both problems mentioned above disappear. $\partial x\bar
\partial x$ is removed from the spectrum, as are all R operators with
$\epsilon=+1$ and the NS operators with $k<0$. We are left with the
operators
\eqn\topspec{T_n=\exp(-\sigma+i(n+{1\over2})x+(n-{1\over2})\phi);\;
V_n=\exp(-{\sigma\over2}-{i\over2}h+inx+(n-1)\phi);\;n=0,1,2,..}
The correlation functions are now all bulk, and we have to divide them
by $\log\mu$. Assuming that the correct prescription to calculate
$N$ point functions is by inserting $N-1$ operators \topspec\
and one conjugate operator (with $k<0$), the amplitudes are very simple
to obtain from the discussion above. At $\mu=0$ we have, e.g.
(after redefining the operators as usual):
\eqn\topcor{\langle T_{n_1}..T_{n_N}\rangle=(N-3)!}
It is amusing that after restricting to \QBRST, all space time fermions
are projected out of the spectrum. The reason is that the Liouville
momentum must satisfy $p_{\rm left}=p_{\rm right}$, which is only
possible \topspec\ in the (NS, NS) and (R,R) sectors. We don't know
whether this observation is more general.
This theory deserves a more detailed examination.
Finally we would like to mention that the conjecture that the model
we are discussing is topological is due to E. Martinec
\ref\emil{E. Martinec, private communication.}.

\bigskip

\newsec{Oscillator states and gravitational degrees of freedom.}

In the previous sections we have shown that the simple scattering pattern
in two dimensional string theory is related to
decoupling of the string states at certain discrete momenta. We start
this section by reviewing their form and then discuss some of their
properties. For simplicity, we restrict to $c=1$ ($\alpha_0=0$).

At values of the momenta $\sqrt{2} k\in Z$, the Virasoro representations
degenerate. Hence the spectrum is richer
\ref\SEI{N. Seiberg, unpublished.},
\PO,
\ref\conespec{B. Lian and G. Zuckerman, Yale preprint YCTP-P18-91 (1991);
S. Mukherji, S. Mukhi and A. Sen, Tata preprint TIFR/TH/91-25 (1991);
P. Bouwknegt, J. McCarthy and K. Pilch, preprint CERN-TH.6162/91 (1991).}.
Parametrizing $k={\rone-\rtwo\over\sqrt2}$ ($\rone, \rtwo\in Z_+$),
we have physical states of the form:
\eqn\Vrs{V_{\rone,\rtwo}^{(\pm)}=\left[\partial^{\rone\rtwo}X+...\right]
\exp\left(i{\rone-\rtwo\over\sqrt2}X+\beta_{\rone,\rtwo}^{(\pm)}\phi\right)}
at level ${1\over2}m^2=\rone\rtwo$. $\beta_{\rone,\rtwo}$ can take as usual
\DDK\ two values: $\beta_{\rone,\rtwo}^{(\pm)}=-\sqrt{2}\pm
{\rone+\rtwo\over\sqrt2}$. In section 2 we used the fact that $V_{\rone,
\rtwo}^{(-)}$ decouple in correlation functions of tachyons. More
precisely, all bulk correlation functions of the form
$\langle V_{\rone,\rtwo}^{(-)}T_{k_1}..T_{k_N}\rangle$ where
$k_1,..,k_{N-1}>0$ are generic and $k_N<0$, vanish.
Here we will sketch the proof of this statement. It is in fact more
convenient to prove vanishing of correlators containing
any number of $V^{(-)}$ and tachyons of generic momenta$(\sqrt{2}k\not\in
Z)$:
\eqn\Vminus{
\langle V_{\rone,\rtwo}^{(-)}...V_{r_{2n-1}, r_{2n}}^{(-)}
T_{k_1}..T_{k_{N-n}}\rangle=0}
inductively in $N(\geq4)$.
First one has to check this for $N=4$: consider $(k_1,k_2>0, k_3<0)$:
$$\int d^2z\langle V_{\rone,\rtwo}^{(-)}(0)T_{k_1}(1)T_{k_2}(\infty)
T_{k_3}(z)\rangle$$
By plugging in the kinematics, one may easily check that the result is a
sum of integrals of the form
$\int d^2zz^n\bar z^m (1-z)^\alpha(1-\bar z)^\beta$ where
$n,m\in Z_+$ and $\alpha,\beta\not\in Z$ (for generic $k_1, k_2$). These
integrals vanish by the standard analytic continuation. Hence,
$\langle V^{(-)} TTT\rangle=0$.
Similarly one checks that
$\langle V^{(-)}V^{(-)} TT\rangle=0$
as well. Now suppose we have shown
\Vminus\ for all $N<N_0$; we want to prove it for $N=N_0$. The strategy
involves as before examining the poles of the integral representation
of \Vminus.
The residues of the poles can be checked by a short calculation
to be given by lower point functions of the form \Vminus\ again, which
vanish by hypothesis. Therefore, the $N=N_0$ point function
\Vminus\ has the property that it has {\it no poles} as a function
of the tachyon momenta $k_i$. As before, one can also estimate the large
$k$ behaviour, and find that this (entire) function
of $k_i$ vanishes at infinity
(for a range of values of the
other $\{k_i\}$). Hence, it is zero everywhere \Vminus.
This concludes the proof of decoupling of $V_{\rone,\rtwo}^{(-)}$.

We would like next to make several comments about this result:

\noindent{}1) Decoupling of states with $\beta<-{Q\over2}$ was advocated
in \NATI, from the point of view of 2D gravity. Our results, while probably
related, are not identical: we proved a statement about bulk amplitudes,
where the Liouville wall, which plays a major role in the considerations
of \NATI, is irrelevant; we used an analytic continuation of the amplitudes
(as a very useful technical tool), which as we saw above is not valid for
generic Liouville amplitudes. Also, the decoupling we find is not complete:
if enough of the tachyons in \Vminus\ are at the discrete momenta
$\sqrt{2} k\in Z$ (in the {\it right} branch), the amplitudes
need not vanish; and tachyons of generic $\beta<-{Q\over2}$ do not decouple.

\noindent{}2) The dynamics of $V_{\rone,\rtwo}^{(-)}$ becomes crucial
in the 2D black hole solution of \WBH. It was shown in \BER\ that this theory
is identical to the $c=1$ model described here with the cosmological
term replaced by $\mu V_{1,1}^{(-)}$. $\mu$ is related to the mass of the black
hole.

\noindent{}3) In a recent paper
\ref\GREGNATI{G. Moore and N. Seiberg, Rutgers preprint RU-91-29 (1991).}
it was shown that the $c=1$ matrix model possesses a large symmetry
algebra, closely related to the discrete states $V_{\rone,\rtwo}$.
Here, on the other hand, we have seen that the simplicity of the amplitudes
is directly due to the decoupling of $V_{\rone,\rtwo}^{(-)}$. The two
observations should be related. A symmetry\foot{The standard matter SU(2)
can not be used since it is not a symmetry for generic radius (e.g.
$R=\infty$).}
would explain e.g.
why decoupling of all $V_{\rone,\rtwo}^{(-)}$ is implied by that
of $V_{\rone,0}^{(-)},
V_{0, \rtwo}^{(-)}$ (tachyons at special momenta).

As explained in section 2, the poles that do appear in the final answer
for the $N$ point functions, correspond to intermediate
states in the $(i,N)$ channel. We can now check which of the states
$V_{\rone,\rtwo}^{(+)}$ \Vrs\ appear in this channel.
Straightforward algebra leads to the conclusion that the pole of
$\Gamma(m_i)$ \redef\ at $m_i=-r_1$ corresponds to the intermediate
state $V_{r_1,r_2=N-3}^{(+)}$. Thus, for given $N$ we see in intermediate
channels all states with $m^2=2r_1(N-3)$, as noted in section 2.
As we vary $N$, we find contributions of all physical states. The reason
why only intermediate states with fixed $r_2$ appear for fixed $N$
is actually purely kinematical:
$\langle T_{k_1}..T_{k_M}V_{r_1,r_2}^{(+)}\rangle$ with all $k_i>0$
(which arise as residues of poles in \corN, see \resN) can only be non zero
if $r_2=M-1$ (by momentum conservation and the resonance condition).

We see that the bulk S -- matrix for the tachyon field describes reasonable
space time physics. The (massless scalar) tachyon field couples
to an infinite set of massive higher spin fields, which are essentially
pure gauge (except at particular momenta). The gauge symmetry of string
theory corresponding to decoupling of BRST commutators is responsible for
the restriction of the (on shell) massive fields to discrete momenta.
However, the simplicity of the results \redef, \rfN\ is due
{\it in addition} to decoupling of half of the remaining states
$V_{\rone,\rtwo}^{(-)}$, which is {\it not explained} by these symmetries.
This implies a further simplification in the dynamics, and in particular
is responsible for the fact that the poles in the
S -- matrix occur as a function
of external momenta alone\foot{We have mainly discussed the S -- matrix
for $(N-1,1)$ kinematics, but one can easily see that the same
decoupling of the `wrong branch' discrete states leads to vanishing
of the bulk amplitudes in all other kinematic regions.}. The solvability
of the matrix model is probably closely related to this phenomenon. One of
the most interesting remaining problems is the realization and implications
of this ``new symmtery'' on the space time equations of motion in two
dimensional string theory.
It appears that the discrete momenta must play a special role in the
space time action. There are several properties of the results, which point
to this, all essentially related to the decoupling of $V_{\rone,\rtwo}^{(-)}$
(which we emphasize is not automatically related
to the fact that by gauge invariance massive physical states occur only
at the above discrete momenta).
In particular, applying
the logic of \CFMP\ to our situation, it seems that a tachyon background
$T_k$ \tach\ which satisfies the linearized equations of motion of the
string (is marginal), also solves the exact non linear equations
of motion (is truly marginal), as long as $\sqrt{2}k\not\in Z$. This
would imply that gravitational back reaction is only possible for discrete
momenta $k$ (this is not a field redefinition invariant statement,
nevertheless, if true, it would be important).

Thus, it is important to understand the dynamics of the operators
$V_{\rone,\rtwo}^{(+)}$. The scattering formulae of section 2 diverge
as the tachyon momentum $k\rightarrow n/\sqrt2$, due to the divergence of
the `leg factors' \redef. One way to interpret this divergence
is to note that by KPZ scaling and \redef, an insertion of $T_k$
into a correlator multiplies it by
\eqn\Omk{\Omega(k)={\Gamma(1-\sqrt{2}|k|)\over\Gamma(\sqrt{2}|k|)}
\mu^{{|k|\over\sqrt2}-1}}
As $k\rightarrow {n+1\over \sqrt2}$, we can
interpret the divergence of \Omk\ as a scaling violation:
\eqn\Omn{\Omega_n\simeq {(-)^n\mu^{n-1\over2}\log\mu\over n!}}
Indeed, the bulk correlation functions considered above have precisely
one insertion of $\log\mu$ corresponding to the unique discrete momentum
tachyon. In general, if more than one momentum goes to $n/\sqrt2$,
there are higher powers of $\log\mu$; of course such powers of $\log\mu$
can occur for any $s$.

In fact, one can convince oneself that the appearance of powers of $\log
\mu$ is a generic property of all $V_{\rone,\rtwo}^{(+)}$. In particular bulk
amplitudes this can be easily verified by factorization of tachyon
bulk amplitudes in appropriate channels. Hence we have in general:
$$\langle V_{\rone,\rtwo}^{(+)}..V_{r_{2n-1},r_{2n}}^{(+)}T_{k_1}
..T_{k_{N-n}}\rangle\propto(\log\mu)^n$$
(for generic $k_1,..k_{N-n}$). One can derive the equivalent of \Omn\
for all $V_{\rone,\rtwo}^{(+)}$; we will not do that here.
Similarly, a natural way to interpret the vanishing of amplitudes
involving $V_{\rone,\rtwo}^{(-)}$ is \Omk\ as factors of $1\over\log\mu$
accompanying each $V^{(-)}$. Therefore, in general we have:
\eqn\disc{\langle V_{\rone,\rtwo}^{(+)}..V_{r_{2n-1},r_{2n}}^{(+)}
V_{s_1,s_2}^{(-)}..V_{s_{2l-1}, s_{2l}}^{(-)}
T_{k_1}
..T_{k_{N}}\rangle\propto(\log\mu)^{n-l}}
Correlators which behave as negative powers of $\log\mu$ are interpreted
as vanishing. From eq. \disc\ one can see precisely the interplay
of $V^{(-)}$ and $V^{(+)}$. For bulk amplitudes, for example, we find
zero if $n\leq l$; this is consistent with all the results described above.

Another (inequivalent) way to define correlation functions of
$V_{\rone,\rtwo}^{(+)}$ is to follow the critical string logic. We
illustrate this procedure with the example of $V_{1,1}^{(+)}=
\partial X\bar\partial X$. The $\log\mu$ divergence discussed above is due
in this case to the fact that turning on $\partial X\bar \partial X$ shifts
the dimensions of the exponentials $T_k$, and we have to compensate
by adjusting the momenta $k_i$. Then inserting $\partial X\bar\partial
X$ into a correlation function \tachcor\ corresponds to $\sum_ik_i
{\partial\over\partial k_i}$. A similar procedure can probably be followed
for all the discrete states.

The world sheet supersymmetric case is again very similar. At momenta
of the form $k={\rone-\rtwo\over2}$, where $\rone,\rtwo\in Z_+$ and
$\rone-\rtwo\in2Z$ corresponding to NS states, while $\rone-\rtwo\in2Z+1$
are in the R sector, we have discrete states at level ${1\over2}m^2=
{1\over2}\rone\rtwo$. Thus in the NS sector the discrete momenta
are $k\in Z$ while for R states it's $k\in Z+{1\over2}$
(in agreement with \tmi, \Rmi). The Liouville dressing takes the form
$\beta_{\rone,\rtwo}^{(\pm)}=-1\pm{\rone+\rtwo\over2}$. As in the bosonic
case, $V_{\rone,\rtwo}^{(-)}$ vanish inside correlation functions of
tachyons \Vminus, and Ramond fields $V$ \Vminus. The derivation is completely
parralel to the one in the bosonic case and we leave it to the reader.

\bigskip

\newsec{Comments.}

There is a large number of open problems related to our work. We will
mention here a few.

1) We do not feel that the issue of states with negative energy $(E=\beta
+{Q\over2}<0)$ is well understood. We have shown here that the bulk
S -- matrix, which is the only part of Liouville correlators which is well
understood, has a sensible interpretation which includes
such states. It is true that the discrete states with $E<0$ partially
decouple, but this is not true for tachyons of generic momentum, and also
breaks down if we turn on discrete states with $E>0$. States with $E<0$
do not correspond to small deformation of the world sheet surface from the
point of view of $2D$ gravity, but they should still play an important
role in the dynamics (e.g. the black hole \WBH, \BER).

2) One would like to have a {\it useful} description of
the space time physics described by the amplitudes we have found --
perhaps a simple action principle for the tachyon and massive degrees
of freedom. As discussed above, this should be different from the
existing string field theories \SFT, \SENG, \GK. In particular,
it would be interesting to incorporate the partial decoupling of $V^{(-)}$
and understand whether there are new symmetries (perhaps related
to those of \GREGNATI,
\ref\W{E. Witten, IAS preprint IASSNS-HEP-91/51 (1991).})
which are responsible for this. Of course,
such a formulation would be useful to study gravitational back reaction
and other issues in this theory.

3) There are extensions and applications of our results
which may be interesting.

\noindent{}a) It is important to derive our results for the extension
from the ``bulk'' to the ``boundary'' correlation functions, which we got
by using space time arguments, directly from the world sheet Liouville
theory. This should shed some light on the origin of the local action
for the tachyon field.

\noindent{}b) We have restricted our attention to genus zero (tree
level) amplitudes. From matrix models \MOORE, \INTER\ we know that
the results for higher genus are almost as simple, and it would be nice
to understand them too from the continuum. We would like to point
out in this context, that one may have problems of convergence
of the appropriate integral representations (which are again trivial
generalizations of the 26 dimensional ones \GSW):
the sum rule $\sum_im_i=1$ \genkin\ (for $s=0$), is replaced
for genus $h$ by
\eqn\sumh{\sum_{i=1}^{N-1}m_i=1-2h}
and since one still expects divergences when $m_i<0$, there is probably
no region where the integral representation converges. The space time
picture should be useful here, as in the spherical case,
and we expect a similar analysis to give the results of \MOORE.

\noindent{}c) It would be interesting to see what properties survive in more
``realistic'' string theories. The natural candidates to consider
are the non critical superstrings \KUS,
where one can increase the number of degrees of freedom in a controlled
way, without losing stability of the vacuum.
We have seen that in two dimensions the theory of \KUS\ is topological.
Its properties should be elucidated further. One may study the
related heterotic theories, which are probably topological as well;
they comprise a large class of theories which are probably completely
solvable.

\noindent{}d) We saw that the 2D fermionic string is described
in space time by
a field theory with two (bosonic) fields, whose tree level S -- matrix
is exactly known. One approach to calculate higher genus corrections would
be to try to write a space time theory similar to the Das- Jevicki
one \SFT, now with two fields; hopefully the tree level structure,
which we have found explicitly, will determine it uniquely. Then one can use
this action in the standard way \INTER\ to get all order results.
This should (among other things) shed light on \DIDK.

\noindent{}e) We have treated here two dimensional strings with
$N=0,1$ SUSY. For $N=2$ two dimensional string theory is critical, and has
been recently shown to possess some interesting features
\ref\Ntwo{H. Ooguri and C. Vafa, Mod. Phys. Lett. {\bf A5} (1990) 1389;
Harvard/Chicago preprint HUTP-91/A003, EFI- 91/05 (1991).}.
We saw that the cases $N=0,1$ give similar space time physics and are
closely related to the $c=1$ matrix model.
The situation is reminiscent of the
relation between the $N=0,1,2$ minimal models of \BPZ, \FR\ in flat
space.
Using our techniques, it is easy to show that all $N\geq4$ point functions
in critical $N=2$ string theory vanish, in agreement with \Ntwo.
The reason is that as emphasized in \Ntwo\ the theory is really
four dimensional, but there is again only one field theoretic
degree of freedom. Unlike the $N=0,1$ cases, here the four dimensional
kinematics implies vanishing of the amplitudes.
It would be interesting
to understand the connection between the work of \Ntwo\ and the theories
described here.

4) One interesting application is to the two dimensional black hole of
\WBH. To understand that, we have \BER\ to replace the cosmological
term $T_{k=0}$ by $V_{1,1}^{(-)}$. First, it is clear that all bulk
amplitudes ($s=0$) of tachyons are the same as in the black hole solution
of \WBH\ and the usual $c=1$ case considered in this paper\foot{Since
all discrete states of the $c=1$ model appear as intermediate states
in such amplitudes (see sections 2,4) we immediately conclude
that all the $c=1$ discrete states must be physical in the black hole
background as well. Furthermore, no additional discrete states
appear in intermediate channels. This suggests that all
other states (e.g. those of \BER) decouple in the bulk, and perhaps
also in general. This phenomenon was demonstrated in \BER.}.
Also, we saw that bulk amplitudes containing tachyons
of generic $k$ and $V_{1,1}^{(-)}$ vanish. We see again that to solve
the black hole theory we must understand the dynamics of the discrete
states $V_{\rone,\rtwo}^{(+)}$, since only they couple to $V^{(-)}$.
The resulting picture of gravitational back reaction in 2D string theory
should be fascinating.

\bigbreak\bigskip\bigskip\centerline{{\bf Acknowledgements}}\nobreak

We would like to thank
M. Bershadsky, P. Freund,
E. Martinec, N. Seiberg, S. Shenker and J.-B. Zuber
for valuable discussions.
D. K. thanks the MSRI, Berkeley and the Aspen Physics Center
for hospitality.
This work was partially supported by
NSF grant PHY-8512793.

%
%
%
%
\vfill
\eject
\appendix{A}{Comparison to KdV.}

The solution of random multimatrix chain-interacting models can
be expressed
\D\ in terms of certain
differential operators $Q=D^n-{n\over2}uD^{n-2}+...$
and $P=D^p-{p\over2}uD^{p-2}+...$,
satifying the
`string equation' \D:
\eqn\streqn{ [P,Q]=1}
The solution of this system of coupled differential equations for the
coefficients of $P$ and $Q$ yields in particular the string
susceptibility $u = \partial_x^2 \log Z=\langle {\cal P}
{\cal P} \rangle$.
In the following we restrict ourselves to the `unitary case'
$deg(P)=n+1$, $deg(Q)=n$, where the
explicit solution of \streqn\ on the sphere is
known \DIFK. The solution is phrased in terms of the
pseudo-differential operator $L$, which satisfies \D,
$L=Q^{1\over n},\;P=L^{n+1}_+$. Operators are defined by
generalized KdV flows:
\eqn\gkdv{
\partial_{t_j} L = [L^j_+,L]  \ \ \ , \ \ \ j=1,2,3,...}
or in terms of the partition sum:
$\partial_{t_j}u=\langle\phi_j{\cal P}{\cal P}\rangle=-2({\rm Res} L^j)^
\prime$. The only feature of the solution of \DIFK\ that we will need
is, that $L$ satisfies (see \DIFK\ for notation and derivations):
\eqn\LJ{L^j_-=-({u\over2})^jD^{-j}+O(D^{-j-1})}
Using \LJ\ it was shown in \DIFK\ that:
\eqn\resdifk{
\eqalign{
\langle \phi_j \phi_m \rangle &= -2 \int Res[L^j_+,L^m_-] = j x^{2 \Delta_j
-\gamma_{str}} \delta_{j,m} \ \ j,m <n. \cr
\langle \phi_j \phi_l \phi_m \rangle &= 2 \int Res([[L^l_-,L^j_+],L^m_-]
-[L^j_+,[L^l_+,L^m_-]) \cr
&=jlm x^{\Delta_j+\Delta_l+\Delta_m -\gamma_{str}-1} N_{jlm} \ \ j,l,m<n. \cr}}
where the scaling dimensions $\Delta_j={{j-1} \over 2n}$ and string
susceptibility exponent $\gamma_{str}= -{1 \over n}$ are the KPZ
exponents \KPZ, \DDK\
for the unitary CFT $(n+1,n)$ coupled to gravity. Note the appearance of the
CFT fusion coefficients $N_{jlm} \in \{ 0,1 \}$ for the three point
functions. The operators $\phi_j$
with $j<n$ were singled out in the calculation:
they correspond to the order parameters of the theory, whose definition is
unambiguous \MSS. The results \resdifk\ agree with \minimal.

In section 2 we also considered $N$ point functions without screening.
For the order parameters, which are the only operators that are simple
to treat using KdV \gkdv\ we have,
$k_p=\alpha_0(1-{j_p}), \ p=1,..,N-1$ and $k_N=\alpha_0(1+
{j_N})$).
Note that $\alpha_0<0$ therefore $k_p>0$; hence we don't have to
worry about the zero energy cuts in $N$ point functions (e.g. \fourpnt),
and the Liouville result we have to compare to is \rfN.
The sum rule \consN\ takes the form:
$j_1+j_2+..+j_{N-1}
=j_N + N-2$. Thus $N$ point functions without screenings correspond
to `the boundary of the fusion rules'. In this case, using \LJ\ it is easy
to calculate all $N$ point functions. Indeed:
\eqn\nptkdv{
\eqalign{
\langle \phi_{j_1} ... \phi_{j_N} \rangle
&=-2\int Res[L^{j_1}_+,[L^{j_2}_+,[..
L^{j_{N-1}}_+,L^{j_N}_-]..] \cr
&=-2 j_1 j_2 ... j_N \int Res(L^{j_1+j_2+..+j_{N-1}-N+1} [L,[L,[..[L,
{{L^{j_N}_-} \over j_N}]..]) \cr
&=j_1 j_2 ... j_N F_N(j_N) \cr}}
The second line results from the fact that we work on the sphere, where each
commutator acts with one derivative only. In \nptkdv\ we have strongly
used eq. \LJ.
Note the close correspondence between \nptkdv\ and the Liouville
calculation: after we factor a product of normalization factors
(which are of course different in the two cases,
compare to \f) we are left with a function of $s$ or $j_N$,
only.
As in the Liouville case, the function of $j_N$, $F$, is now determined
by putting $N-3$ of the $j_i$ to 1. Then we can use the result
for the three point function \resdifk, to find
$F_N= (\partial_x)^{N-3} x^{s+N-3}$, where $s=\sum_{p=1}^N \Delta_{j_p}
-\gamma_{str}-N+2$, the correct KPZ scaling for the $N$ point function,
and finally:
\eqn\resnkdv{ \langle \phi_{j_1} ... \phi_{j_N} \rangle = j_1 j_2 .. j_N
(\partial_x)^{N-3} x^{s+N-3} }
In agreement
(up to a different normalization of the
operators) with the Liouville result \rfN.

%
%
%
%
%
%
%
%
%
%
%

\appendix{B}{1PI calculus.}

This appendix is devoted to various calculations of 1PI vertices at
$c=1$ ($\alpha_0=0$). In sect. 2.2.6, we have shown how to compute the
general 1PI vertices $A_{1PI}^{(N)}(k_1,..,k_N)$ directly:
it is the sum over all tree
graphs with $N$ external legs carrying the momenta $k_1,..,
k_{N-1}>0$, $k_N<0$,
and the following Feynman rules:

\noindent{}1) propagators: $-{{\vert k \vert} \over
\sqrt{2}}$ for each internal leg carrying the total momentum $k$
(momentum is conserved at the vertices).

\noindent{}2) vertices: ${\tilde A}(l_1,..,l_n)=(\partial_{\mu})^{n-3}
{\mu}^{\sqrt{2}\vert l \vert -1} \vert_{\mu=1}$ for each $n$-legged
vertex with incoming momenta $l_1$,..,$l_n$, $l$ denoting the only negative
momentum among these.

To illustrate the procedure, let us calculate $A_{1PI}^{(4)}$
again, using the new
method: there are four trees with external momenta $k_1$,..,$k_4$, the
$s$, $t$, $u$ channels and the maximal star of one 4-legged vertex.
Adding up the four contributions we find:
\eqn\fouripione{
\eqalign{
A_{1PI}^{(4)}(k_1,..,k_4)&= -{1 \over \sqrt{2}}\big[ \vert k_1+k_2 \vert
+ \vert k_1+k_3 \vert+ \vert k_2+k_3 \vert \big] + (\sqrt{2} \vert k_4
\vert -1) \cr
&= -1 \cr}}
where obvious use of the conservation law $-k_4=k_1+k_2+k_3$ has been made.

Repeating the same procedure for $N=5,6$ yields:
\eqn\reipifs{
\eqalign{
A_{1PI}^{(5)}&= 2 - {1 \over 2} \sum_{i=1}^5 k_i^2 \cr
A_{1PI}^{(6)}&= -6 + 3 \sum_{i=1}^6 k_i^2 \cr}}
Note that the irreducible vertices are no longer constants.
The main problem with these computations
is that they involve writing all tree graphs with $N$ external legs
whose number grows very quickly ($26$ in the case $N=5$, $236$ in the case
$N=6$). We will present below a simple recursive way of generating
arbitrary 1PI vertices.

The first simple object one can look at is the vertex with, say $p$ non-zero
momenta and $N-p$ zero momenta $A_{1PI}^{(N)}(k_1,..,k_p,0,..,0)$.
Using the method described in the begining of this appendix,
it is easy to see the effect of
adding one zero-momentum external leg to such a vertex: due to the form of
the propagator $\pi(k)= -{\sqrt{2} \over 2} \vert k \vert$, the only non-zero
contributions to the sum over trees come from either an addition
on a leg carrying a non-zero momentum $k$ (multiplication by $-{\sqrt{2}
\over 2} \vert k \vert$), or an addition on a vertex $V_n(k)=
(\partial_{\mu})^{n-3} \mu^{\sqrt{2}\vert k \vert -1} \vert_{\mu=1}$,
which simply changes it into $V_{n+1}$.
By recursion, it is straightforward to show that:
\eqn\pinp{
\eqalign{
A_{1PI}^{(N)}(k_1,..,k_p,0,..,0)&= (\partial_{\mu})^{N-p}
\prod_{i=1}^p {2 \over {1+\mu^{\sqrt{2}\vert k_i \vert}}} \cr
&\sum_{trees(k_1,..,k_p)} (\pi(k)= -{\sqrt{2}\vert k \vert \over
{1 + \mu^{\sqrt{2} \vert k \vert }}}; V_n(k)=(\partial_{\mu})^{n-3}
\mu^{\sqrt{2}\vert k \vert-1})
\bigg\vert_{\mu=1} \cr}}
where the sum extends to all trees with external momenta $k_1>0$,...$k_{p-1}>0$
and $k_p <0$; the notation $(\pi(k)=.. ; V_n(k)=..)$ means that a weight
$\pi(k)$ has to be attached to each internal leg carrying the momentum
$k$, and a weight $V_n(k)$ has to be attached to each $n$-legged vertex
whose only negative external momentum is $k$.
Note that the external legs receive a weight ${2 / {1 + \mu^{\sqrt{2} \vert
k \vert }}}$.
It is an easy exercise to see that
a differentiation w.r.t. $\mu$ exactly reproduces the above additions.

As an example, in the case $p=2$, eqn.\pinp\ yields:
\eqn\pitwo{
\eqalign{
A_{1PI}^{(N)}(k,-k,0,..,0)&= -{\sqrt{2} \over
{\vert k \vert}} (\partial_{\mu})^{N-2}
{2 \over {1+\mu^{\sqrt{2} \vert k \vert}}} \bigg\vert_{\mu=1} \cr
&=
-{\sqrt{2} \over \vert k \vert} (\partial_{\mu})^{N-2}
(1 - \tanh({\vert k \vert \over \sqrt{2}}\log \mu)) \bigg\vert_{\mu=1} \cr}}
from which we get immediately:
\eqn\ripin{
\eqalign{
A_{1PI}^{(3)}&=1 \cr
A_{1PI}^{(4)}&=-1 \cr
A_{1PI}^{(5)}&= 2 -k^2 \cr
A_{1PI}^{(6)}&= -6 + 6 k^2 \cr
A_{1PI}^{(7)}&= 24 -35 k^2 +4 k^4 \cr
A_{1PI}^{(8)}&= -120 +225 k^2 -60 k^4 \cr
A_{1PI}^{(9)}&= 720 -1624 k^2 + 700 k^4 -34 k^6 \cr}}

In the case $p=3$, \pinp\ is still very simple because the sum reduces to only
one term, with weight $\mu^{\sqrt{2}\vert k_3 \vert}$, so that:
\eqn\thrbett{
A_{1PI}^{(N)}(k_1,k_2,k_3,0,..,0)=(\partial_{\mu})^{N-3}\left\{
\mu^{\sqrt{2} \vert k_3 \vert} \prod_{i=1}^3 {2 \over {1 + \mu^{\sqrt{2}
\vert k_i \vert}}} \right\}\bigg\vert_{\mu=1} }
or, by redistributing the power $\sqrt{2} \vert k_3 \vert = {1 \over \sqrt{2}}
(\vert k_1 \vert+ \vert k_2 \vert + \vert k_3 \vert)$ onto the individual
leg factors, this can be put in the form \threenz.

In fact, the general expression \pinp\ can be improved as follows:
\eqn\pimpr{
\eqalign{
A_{1PI}^{(N)}(k_1,..,k_p,0,..,0) &= (\partial_{\mu})^{N-p} \prod_{i=1}^p
{1 \over {\cosh({k_i \over \sqrt{2}}\log \mu)}} \cr
&\sum_{trees(k_1,..,k_p)}
(\pi(k)= -\mu \partial_{\mu} \log \cosh({k \over \sqrt{2}}\log \mu);
V_n= \mu^{2-n} A_{1PI}^{(n)}) \bigg\vert_{\mu=1} \cr}}
which yields \threenz, \fournz\ in the particular cases $p=3,4$.
To get \pimpr\ from \pinp,
we reabsorbed a factor $\mu^{\vert k \vert \over \sqrt{2}}$ into each leg
around a vertex, yielding the product of external leg weights prefactor,
and a propagator
$$  \pi(k)=
-\mu \partial_{\mu} \log(1+\mu^{\sqrt{2} \vert k \vert})
= -{\vert k \vert \over \sqrt{2}} -\mu \partial_{\mu}
\log \cosh({k \over \sqrt{2}} \log \mu),$$
and performed the partial sums corresponding to the $-{\vert k \vert \over
\sqrt{2}}$
piece of the propagator, yielding the vertices $V_n= \mu^{2-n} A_{1PI}^{(n)}$.

In the case $p=N-1$, the expression \pimpr\ gives rise to a very simple
recursion relation:
\eqn\finrec{
\eqalign{
A_{1PI}^{(N)}(0,k_1,..,k_{N-1})&= (3-N)A_{1PI}^{(N-1)}(k_1,..,k_{N-1}) - \cr
&-\sum_{2 \leq p < {N \over 2};\sigma } {l^2 \over 2}
A_{1PI}^{(p+1)}(k_{\sigma(1)},..,k_{\sigma(p)},l)
A_{1PI}^{(N-p)}(l,k_{\sigma(p+1)},..,k_{\sigma(N-1)}) \cr}}
where for each $p$ the sum extends over the permutations $\sigma$
of $\{ 1,..,N-1 \}$ yielding distinct sets $\{ \sigma(1),..,\sigma(p) \}$
(the symmetric term $N-p=p+1$ is counted only once), and $l$ denotes
the intermediary momentum fixed by the conservation law.
This expression shows explicitly
that $A_{1PI}^{(N)}$ with one zero external momentum is a polynomial
in the variables $k_{I_j}^2=(\sum_{i \in I_j} k_i)^2$, $I_{j} \subset
\{ 1,..,N-1\}$, with total degree $N-4+(N \ mod \ 2)$.
The general vertex is then obtained by symmetrization
of \finrec\ w.r.t. $k_N$.
As an example we quote the case $N=7$:
\eqn\seven{
A_{1PI}^{(7)} = 24 -{35 \over 2} \sum_{i=1}^7 k_i^2
+ (\sum_{i=1}^7 k_i^2)^2 + {1 \over 4} \sum_{1 \leq i < j \leq 7}
(k_i + k_j)^2 [(k_i+k_j)^2 -k_i^2-k_j^2], }
valid for all momenta.

\listrefs
\end